\documentclass[onecolumn,secnumarabic,amssymb, nobibnotes, aps, prb]{revtex4-2}
\usepackage{graphicx}
\usepackage{caption}
\usepackage{subcaption}
\usepackage{xcolor}
\newcommand{\pmb}[1]{{\setbox0=\hbox{#1}%
  \kern-.025em\copy0\kern-\wd0
  \kern.05em\copy0\kern-\wd0
  \kern-.025em\raise.0433em\box0 }}
\newcommand{\fg}[1]{\mbox{\pmb{$#1$}}}
\newcommand{\bey}{\begin{eqnarray}}
\newcommand{\eey}{\end{eqnarray}}

\newcommand{\sg}{\sigma}

\newcommand{\env}[1]{\texttt{#1}}
\begin{document}

\title{Effect of initial microstructure on {\color{black} its  evolution and $\alpha \rightarrow \omega$ phase transition in Zr under hydrostatic loading}}%

\author{K. K. Pandey}%
\email{kkpandey@barc.gov.in}
\affiliation{High Pressure \& Synchrotron Radiation Physics Division, Bhabha Atomic Research Centre, Trombay, Mumbai 400 085, India}
\affiliation{Homi Bhabha National Institute, Anushaktinagar,  Mumbai 400 094, India}
\author{Valery I. Levitas}
\email[Corresponding author: ]{vlevitas@iastate.edu}
\affiliation{Department of Aerospace Engineering, Iowa State University, Ames, Iowa 50011, USA}
\affiliation{Department of Mechanical Engineering, Iowa State University, Ames, Iowa 50011, USA}
\affiliation{Ames Laboratory, U.S. Department of Energy, Iowa State University, Ames, Iowa 50011-3020, USA}
\author{Changyong Park}
\affiliation{High Pressure Collaborative Access Team, X-ray Science Division, Argonne National Laboratory, Argonne, IL, 60439 USA}
\author{Guoyin Shen}
\affiliation{High Pressure Collaborative Access Team, X-ray Science Division, Argonne National Laboratory, Argonne, IL, 60439 USA}

\date{\today}%
\begin{abstract}
 {\color{black} The first study of the effect of the initial microstructure on its evolution under hydrostatic compression before, during, and after the irreversible $\alpha\rightarrow\omega$ phase transformation and during pressure release} in  Zr using in situ x-ray diffraction is presented. Two samples were studied:
 one is plastically pre-deformed Zr with saturated hardness and the other is annealed.
{\color{black}  Phase transformation $\alpha\rightarrow\omega$  initiates at lower pressure for pre-deformed sample
but above volume fraction of $\omega$ Zr $c= 0.7$, larger volume fraction is observed for the annealed sample. 
This implies that the general theory based on the proportionality between the
athermal resistance to the transformation and the yield strength must be essentially advanced.  }
The crystal domain size significantly reduces, and microstrain and dislocation density increase during loading for both  $\alpha$ and $\omega$ phases in their single-phase regions. For the $\alpha$ phase,
domain sizes are much smaller for prestrained Zr, while microstrain and dislocation densities
are much higher. {\color{black} For the cold-rolled sample at 5.9 GPa (just before initiation of transformation), domain size in $\alpha$ Zr
decreased to $\sim 45\, nm$ and dislocation density increased to $1.1 \times 10^{15}\, lines/m^2$, values similar to those after severe
plastic deformation under high pressure. Despite the generally accepted concept that hydrostatic pressure does not cause plastic straining,
it does and is estimated.
During transformation, the first rule was found: The average domain size, microstrain, and dislocation density in  $\omega$ Zr for $c<0.8$ are functions of the volume fraction of $\omega$ Zr only, which
  are independent of the plastic strain tensor prior to transformation and pressure.
The microstructure is not inherited during phase transformation.
Surprisingly,
for the annealed sample, the final dislocation density and average microstrain after pressure release in the $\omega$ phase are larger than for the severely pre-deformed sample.
 The significant evolution of the microstructure and its effect on phase transformation demonstrates that their postmortem evaluation does not represent the actual conditions during loading.}
 A simple model for the initiation of the phase transformation involving microstrain is suggested. The results suggest that an extended experimental basis is required for the predictive models for the combined pressure-induced phase transformations and microstructure evolutions.
\end{abstract}

\maketitle
\section{Introduction}

{\color{black} It is generally accepted, starting with classical Bridgman's work \cite{Bridgman-52}, and is used in all plasticity books \cite{lubliner-1990,levitas-book96} that hydrostatic loading of void-free materials does not cause plastic deformation.
However, even within liquid or gas pressure transmitting media,  various defects, like dislocations, twins, grain boundaries, and their junctions, cause essential local internal stresses with deviatoric/shear stress components, which may cause plastic deformations. Change in volume and shape during phase transformation causes strong internal stresses leading to significant plasticity, which in turn affects the phase transformation progress. Traditionally, plasticity is characterized in terms of residual (plastic strain), but under hydrostatic conditions, the shape of a macroscopic sample should not change, thus, plasticity is not measurable. However,
the width of the x-ray diffraction (XRD) peaks broadens, and we can extract from this broadening the dislocation density, domain size, and microstrain, and utilize them for physical (instead of geometric) in situ characterization of plasticity under hydrostatic conditions.
We are not aware of similar efforts in the existing literature.
}

{\color{black}  A special industry produces nano-grained materials by severe plastic deformation under high pressures, and corresponding basic science studies the evolution of microstructural parameters \cite{edalati2022nanomaterials}.
However, the individual role of plastic straining and pressure is not clear.
}

It is well known that plastic deformation strongly affects phase transformations in various materials 
\cite{levitas-mechchem-04,levitas-prb-04,Blank-Estrin-2014,Edalati-Horita-16,Gaoetal-19,Levitas-JPCM-18,Levitas-MT-19,Levitas-MT-23}, however, its general understanding and quantitative descriptions are still lacking.
For example,  the well-known hcp ($\alpha$) to simple hexagonal ($\omega$) phase transformation of Zr has been reported over a broad pressure range of {\color{black} 0.67} - 7 GPa \cite{Zilbershtein-75,Banerjee-2007,Perez-2009,Edalatietal-MSEA-2009,Zhilyaevetal-MSEA-2011,Blank-Estrin-2014,KKP-Levitas-Acta-2020,Lin-Levitas-MRL-23},
which shows largely scattered data from various researchers, some results are even contradictory.
{\color{black} In comparison, under hydrostatic loading, this PT occurs in the pressure range of 5 - 17 GPa \cite{Velisavljevic-etal-11,Liu-etal-23,Anzellini-etal-20,Kumaretal-Acta-20}.}
It is also found for various materials, including Zr, Ti, and Fe, that plastic straining (e.g., by hydroextrusion) prior to transformation increases pressure hysteresis under hydrostatic loading (defined as the difference between pressures for the initiation of the direct and reverse transformations) \cite{Blank-Estrin-2014}. However, it is also reported that if transformation occurs during the plastic straining (e.g., during plastic shearing at high pressures),  then the pressure hysteresis for  $\alpha \rightarrow \omega$ phase transformation in Zr and Ti reduces down to zero \cite{Zilbershtein-75,Blank-Estrin-2014}.

The early studies of the effect of plastic strain on phase transformations recognized that there are different types of phase transformation under high pressure  \cite{levitas-mechchem-04,levitas-prb-04}, namely pressure-, stress-, and plastic strain-induced phase transformations. Both
{\it pressure-induced phase transformations} under hydrostatic loading and {\it stress-induced phase transformations} under non-hydrostatic loading but below the macroscopic yield strength initiate at pre-existing defects in the sample (e.g., dislocations and various tilt boundaries), which represent stress (pressure) concentrators. This implies that the initiation of phase transformation should depend {\color{black} (in addition to the concentration of impurities \cite{Velisavljevic-etal-11})} on the initial microstructure of the sample. However, initial microstructure is seldom {\color{black} \cite{Levitas-JPCM-18,Kumaretal-Acta-20}} characterized or reported in high-pressure studies. This could be one of the reasons for the scattered data in the reported transformation pressure for many materials (including Zr) by different researchers. Besides,  microstructure itself may evolve even under hydrostatic pressures before, during, and after the phase transformations {\color{black} which to the best of our knowledge was never studied in situ in literature. This should affect the PT's progress.}

 {\it Plastic strain-induced phase transformations} under high pressure occur during plastic flow by nucleation at new defects generated during plastic flow \cite{levitas-mechchem-04,levitas-prb-04}. They require completely different experimental characterization and thermodynamic and kinetic treatment, e.g. as described in the four-scale theoretical approaches \cite{Levitas-MT-19} and the most advanced experimental measurements with rotational diamond anvil cell (DAC) on $\alpha \rightarrow \omega$ phase transformation in Zr \cite{KKP-Levitas-Acta-2020}. Still, the kinetic equation for the strain-induced phase transformations contains as a parameter phase transformation pressure under hydrostatic conditions \cite{levitas-mechchem-04,levitas-prb-04,Levitas-MT-19}. Currently, the pressure for initiation of transformation under hydrostatic conditions is used in this equation for describing experimental data \cite{KKP-Levitas-Acta-2020}, while it depends on the volume fraction of the high-pressure phase $c$, microstructure, and plastic strain. This kinetics has been utilized in the macroscopic finite element analysis of strain-induced phase transformations in
rotational DAC \cite{Levitas-Zarechnyy-PRB-DAC-10,Levitas-Zarechnyy-PRB-RDAC-10,Feng-Levitas-IJP-BN-RDAC-19}.
 Thus, knowledge of the evolution of
 kinetics of volume fraction $c$ and microstructural parameters under hydrostatic conditions with pressure $p$ is important for the description of the strain-induced phase transformations as well.

 There is a contradiction in understanding the effect of pre-existing plastic deformation on pressure-induced transformation. As we mentioned, experiments in  \cite{Blank-Estrin-2014} report proportionality between
 the pressure hysteresis and material hardness, which is varied by varying plastic strain prior to transformation. Since plastic deformation increases the yield strength and hardness (which are related by a coefficient in the range from 0.333 to 0.386) with the plastic strain, it should increase pressure hysteresis.
Assuming equal deviation of the pressures for direct and reverse transformations from the phase equilibrium pressure,
this means that the deviation 
increases with increasing plastic strain. 
{\color{black} This result was formalized in the theory  for pressure- and stress-induced phase transformations in
\cite{Levitas-JMPS-I-97,Levitas-JMPS-II-97,levitas-ijss1998,Levitas-IJP-21}
in the form $X=k$ and $k=L \sg_y$, where $X$ is the thermodynamic driving force for the phase transformation,
$k>0$ is the athermal resistance to the transformation (dissipative threshold), which is proportional to the yield strength with a proportionality factor $L$.
It was explicitly  used in
\cite{Levitas-JMPS-I-97,Levitas-JMPS-II-97,levitas-ijss1998,Levitas-IJP-21}  in the analytical solutions
and in \cite{Levitasetal-PhilMag-02} for the finite element simulations of various phase transformation problems.}
However, it is also  known that dislocations serve as nucleation sites for the initiation of phase transformation
 \cite{olson+cohen-1986,levitas-mechchem-04,levitas-prb-04}, which is confirmed by more recent
 phase-field simulations \cite{Xu-Khachaturyan-etal-NPJ-CM-18,Xu-Khachaturyan-etal-AM-19,levitas-javanbakht-APL-13,Javanbakht-Levitas-JMS-18}.
 Consequently, an increase in plastic strain increases the number of dislocations and promotes nucleation.
 Recent experiments \cite{Kumaretal-Acta-20} on $\alpha \rightarrow \omega$ phase transformation in Zr show that  plastic compression by $5\%$ and $10\%$ prior to transformation shifts the entire kinetic curve 
 to lower pressures with increasing plastic strain which contradicts the expected increase in pressure hysteresis according to  \cite{Blank-Estrin-2014}. This was explained by the increasing role of stress concentrators due to increasing activation of twinning in $\alpha$ Zr during the plastic compression.


{\color{black} One more general question is whether microstructural parameters measured post-mortem after pressure release represent
those under high pressure. To address it, one needs, in particular, to study microstructure evolution during pressure release.
}

  With these motivations, {\color{black} we performed here the first in situ XRD study of the effect of the initial microstructure (domain size, microstrain, and dislocation density) produced by severe plastic deformation and annealing, respectively, on its evolution before, during, and after the $\alpha\rightarrow\omega$ phase transformation in Zr under hydrostatic compression and during pressure release}.  We follow the program presented in the viewpoint article \cite{Levitas-JPCM-18} and focus on two initial states: one is severely pre-deformed up to maximum hardness, which does not change with further plastic deformation, and the other is annealed at $600^\circ C$.
  Note that studying materials with maximum saturated hardness significantly simplified
  understanding of strain-induced phase transformations  \cite{KKP-Levitas-Acta-2020}.

 \begin{figure}[h!]
\includegraphics[width=0.6\linewidth]{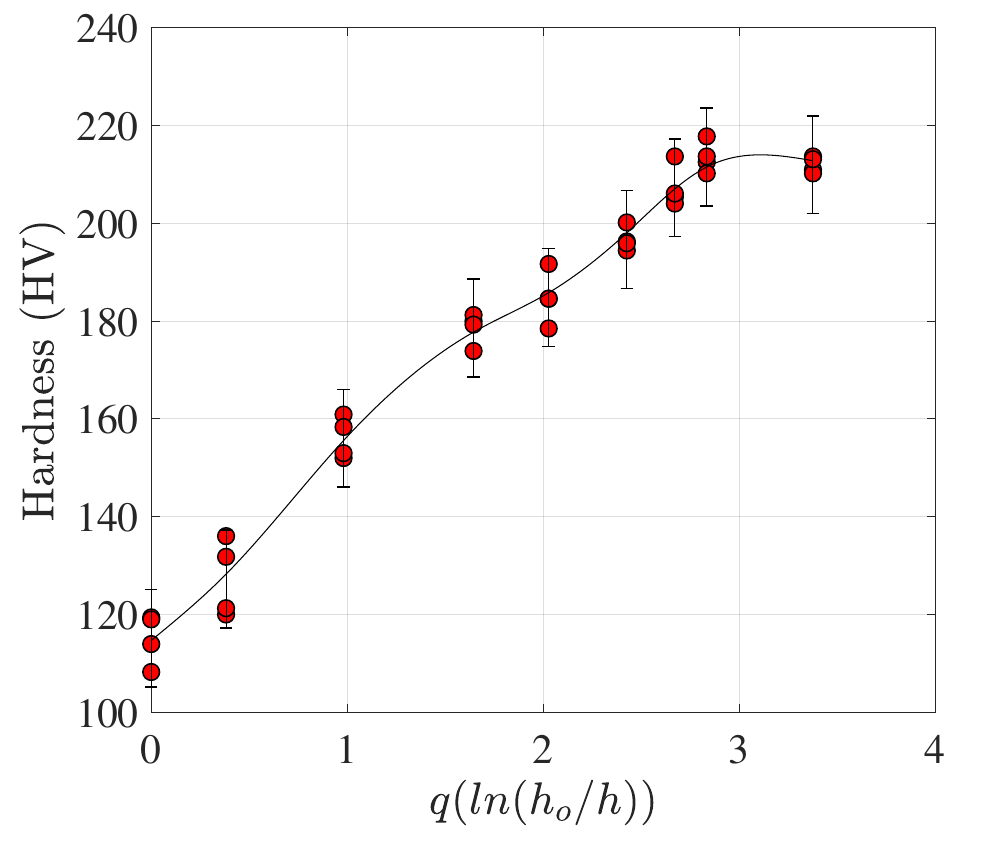}
\caption{Vickers hardness of cold-rolled Zr sample as a function of sample thickness. }
\label{fig:hardness1}
\end{figure}

\begin{figure}[h!]
\includegraphics[width=0.6\linewidth]{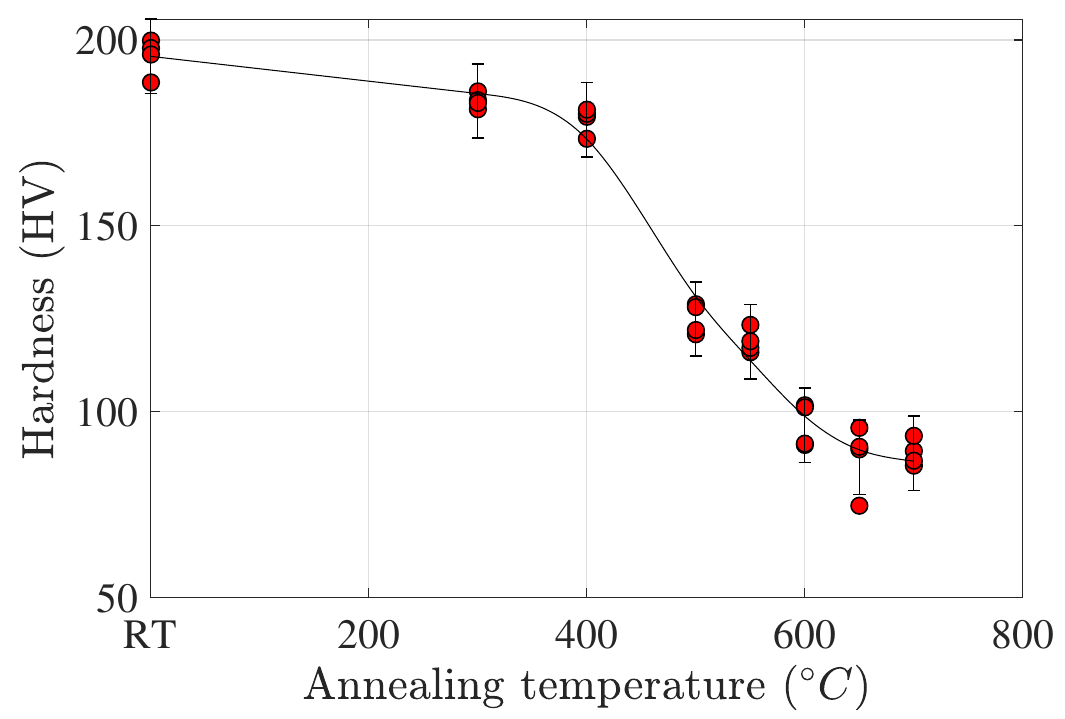}
\caption{Vickers hardness of annealed Zr samples as a function of annealing temperature.}
\label{fig:hardness2}
\end{figure}
\section{Sample preparation}
The material studied here is commercially pure (99.8\%) $\alpha$ Zr (Fe: 330 ppm; Mn: 27 ppm; Hf: 452 ppm; S: $< 550$ ppm; Nd: $< 500$ ppm) with an equiaxed grain size of $13\, \mu m$. Zr samples with varying initial microstructures were prepared from cold-rolled pure-Zr samples plastically deformed to saturated hardness.
To achieve saturated hardness, the initial sample (in the form of plate $\sim 5.2$ mm thick) was cold-rolled in several steps, and at each step, its Vickers hardness \cite{Vickers} was measured using LECO LM 247AT micro-indentation hardness tester at Metallography Laboratory at Iowa State University. The saturated hardness of $\sim 200 HV$ was achieved at a final thickness of $\sim 300\, \mu m$ (Fig. \ref{fig:hardness1}). Several small pieces ($\sim1\times1$ cm) were then cut from this cold-rolled thin sheet. Also, the powdered sample was prepared by the diamond filing of the sheet and subsequent grinding in mortal pastel for about 2 hours.  These small pieces of sheets along with some amount of powdered Zr sample were then annealed in several batches to different annealing temperatures ranging from $300^\circ C$ to $750^\circ C$ in an inert (Ar) environment and subsequently cooled down to ambient temperature at a rate of 100$^\circ C$ per hour for each sample set. After annealing treatment, no further material processing was done on the powdered sample and small pieces of sheets to avoid any further changes in the microstructure and retain the same microstructure in the sheet and the powdered samples in each sample set.

\section{Experimental Methods}
\subsection{Characterization of initial microstructure} To characterize the initial microstructure of each sample set, Vickers hardness measurements were carried out on the small pieces, and XRD measurements were carried out on the powdered sample in each sample set. As shown in Fig. \ref{fig:hardness2}, Vickers hardness drastically reduces above  $400^\circ C$ annealing temperature and reaches the lowest hardness at annealing temperature above $600^\circ C$.

The XRD measurements were carried out on the powdered Zr samples from each sample set using monochromatic X-rays of wavelength $0.3093\AA$ at the bending magnet beamline 16-BM-D at Advanced Photon Source at Argonne National Laboratory, USA. The measurements were carried out in transmission geometry using a focused X-ray beam of size $\sim 7\, \mu m \times 6\, \mu m$. Two-dimensional diffraction images were collected at Perkin Elmer flat panel detector and were converted to one-dimensional diffraction pattern using FIT2D software \cite{fit2d1,fit2d2} and subsequently analyzed using GSAS-II \cite{gsas2} and MAUD \cite{MAUD} software. For sample-detector distance calibration and deconvolution of instrumental broadening, XRD data was recorded on the NIST standard $CeO_2$ sample.
\begin{figure}[h!]
\includegraphics[width=0.8\linewidth]{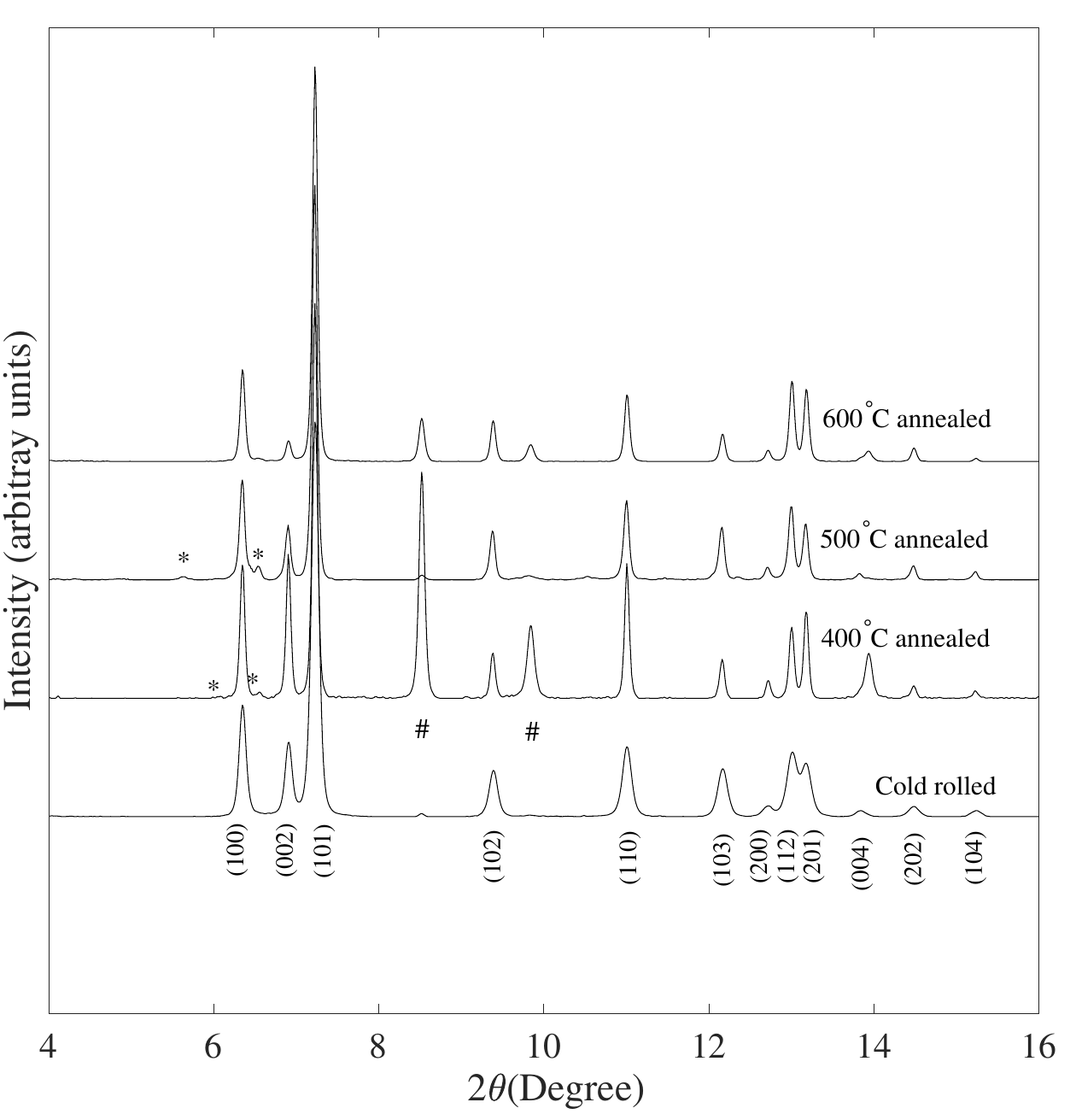}
\caption{X-ray diffraction patterns of cold rolled sample and samples annealed to different temperatures. $\ast$ markers show the diffraction peaks from impurities that appeared during the annealing process and $\#$ mark indicates the (111) and (200) diffraction peaks of Cu. }
\label{fig:ambientstack}
\end{figure}
X-ray diffraction patterns of cold rolled and annealed samples are shown in Fig. \ref{fig:ambientstack}. The samples annealed to $400^\circ C$ and $500^\circ C$ exhibit some impurity peaks after annealing. These diffraction patterns were analyzed to estimate the average crystalline domain size, microstrain, and dislocation density for each sample.  Details of microstructural analysis using XRD and estimation of dislocation density are described in Appendix \ref{microstructureanalysis}.  As shown in Fig. \ref{fig:xrdchar},  the average crystalline domain size increases from $\sim60\, nm$ to a saturated value of $\sim300\, nm$ at annealing temperatures above $400^\circ C$. The average microstrain reduces nearly by an order of magnitude from 0.0031 for the cold-rolled sample to 0.00049 for the annealed sample at $600^\circ C$. Estimated dislocation density also reduces from $\sim 6\times 10^{14}$  $lines/m^2$ to $\sim 1.7\times 10^{13}$  $lines/m^2$.

There is some inconsistency between the hardness test studies and the microstructural results as we observed a significant change in average crystalline domain size and dislocation density for the sample annealed at $400^\circ C$, whereas the hardness of the sample annealed at the same temperature is comparable to that of the cold rolled sample. The XRD measurements were carried out on powdered annealed samples, whereas hardness testing was carried out on thin sheets of samples annealed at the same temperature. The powdered samples with a relatively larger surface area are more prone to impurities during annealing, which is also evident from the XRD pattern, which might have led to the discrepancy. The cold-rolled sample and the sample annealed at $600^\circ  C$, however, exhibit pure $\alpha$ Zr phase. So, high-pressure studies were carried out on these samples only.

\subsection{High-pressure studies} To study the effect of initial microstructure and its evolution on the high-pressure phase transition in Zr, we carried out high-pressure hydrostatic compression experiments on the end member samples, i. e., the cold-rolled sample, and the $600^\circ C$ annealed sample. Silicone oil was used as a pressure-transmitting medium and Cu was used as a pressure marker. Silicone oil is as good as methanol: ethanol mixture for low pressure \textless 20 GPa and even better than methanol: ethanol mixture above 20 GPa \cite{Yongrong-RSI2004}. Solidification of silicone oil under high pressures is not precisely reported as the process could be gradual with pressure in this case \cite{Torikachvili-RSI2015}; though Klotz et al. \cite{Klotz-2009} have shown a slight increase in standard deviation in pressure in DAC  at 2.5 GPa which reduced at further higher pressures, which could be a signature of solidification. Torikachvili et al. \cite{Torikachvili-RSI2015} have shown that even methanol: ethanol mixture solidifies at ~ 6 GPa but is considered hydrostatic up to 12 GPa \cite{Klotz-2009}. In the pressure range of our study, non-hydrostatic stresses in silicone oil are negligible. Silicone oil can also be relatively easily loaded in diamond anvil cells (DAC) due to non-volatility.  For high-pressure experiments, a sample chamber was prepared by drilling a hole of diameter $\sim 250\, \mu m$ at the center of a pre-indented steel gasket thinned down from $\sim 250\, \mu m$ to $\sim 60\, \mu m$. The culet size of diamond anvils used in high-pressure cells was $\sim 500\, \mu m$. The sample was loaded along with a pressure marker and the pressure transmitting medium in the sample chamber. Pressure at the sample was estimated using the well-known equation of state of Cu \cite{Dewaele-PRB-2004}. Experiments were performed in our rotational DAC
\cite{RDAC-ISM-15}  in compression mode because using motorized loading in a rotational DAC system allowed
much smaller controllable load steps than membrane systems for traditional DACs at
beamline 16-BM-D. Thus, XRD images were recorded in pressure steps of ~ 0.2 GPa to have sufficient data points across the  $\alpha\rightarrow\omega$ phase transformation in Zr.
\begin{figure}
\includegraphics[width=\textwidth]{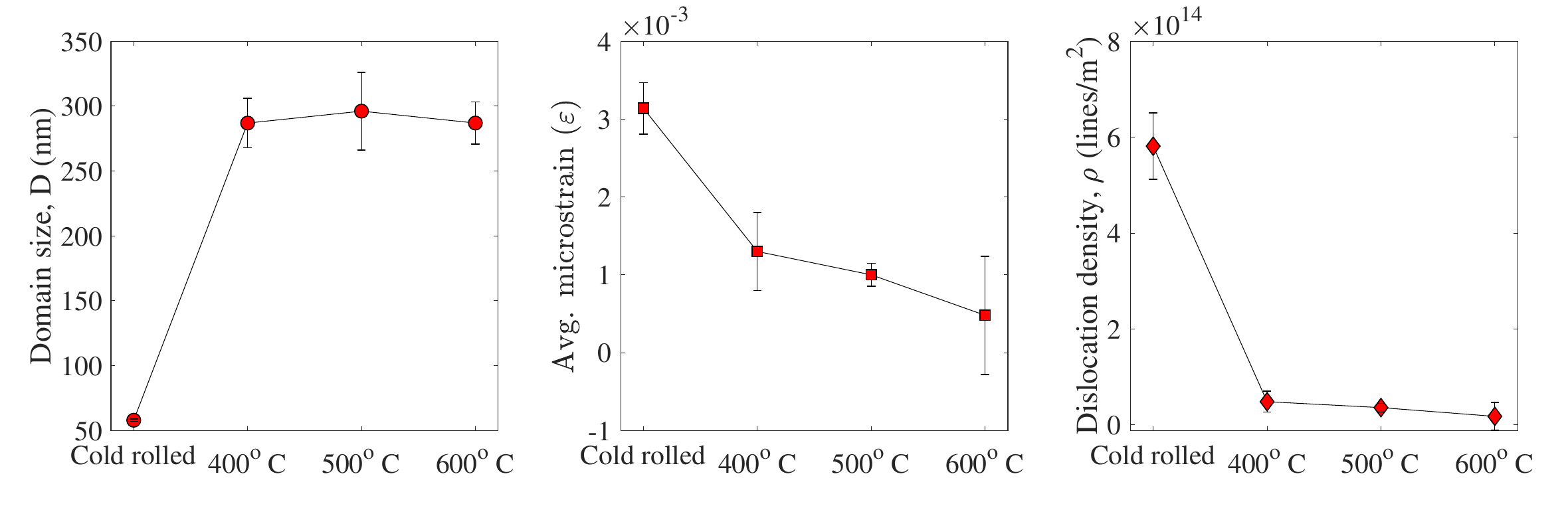}
\caption{Average crystalline domain size, microstrain, and estimated dislocation density as a function of annealing temperature.}
\label{fig:xrdchar}
\end{figure}

\section{Results and discussion}
\begin{figure}[h!]
\includegraphics[width=0.8\linewidth]{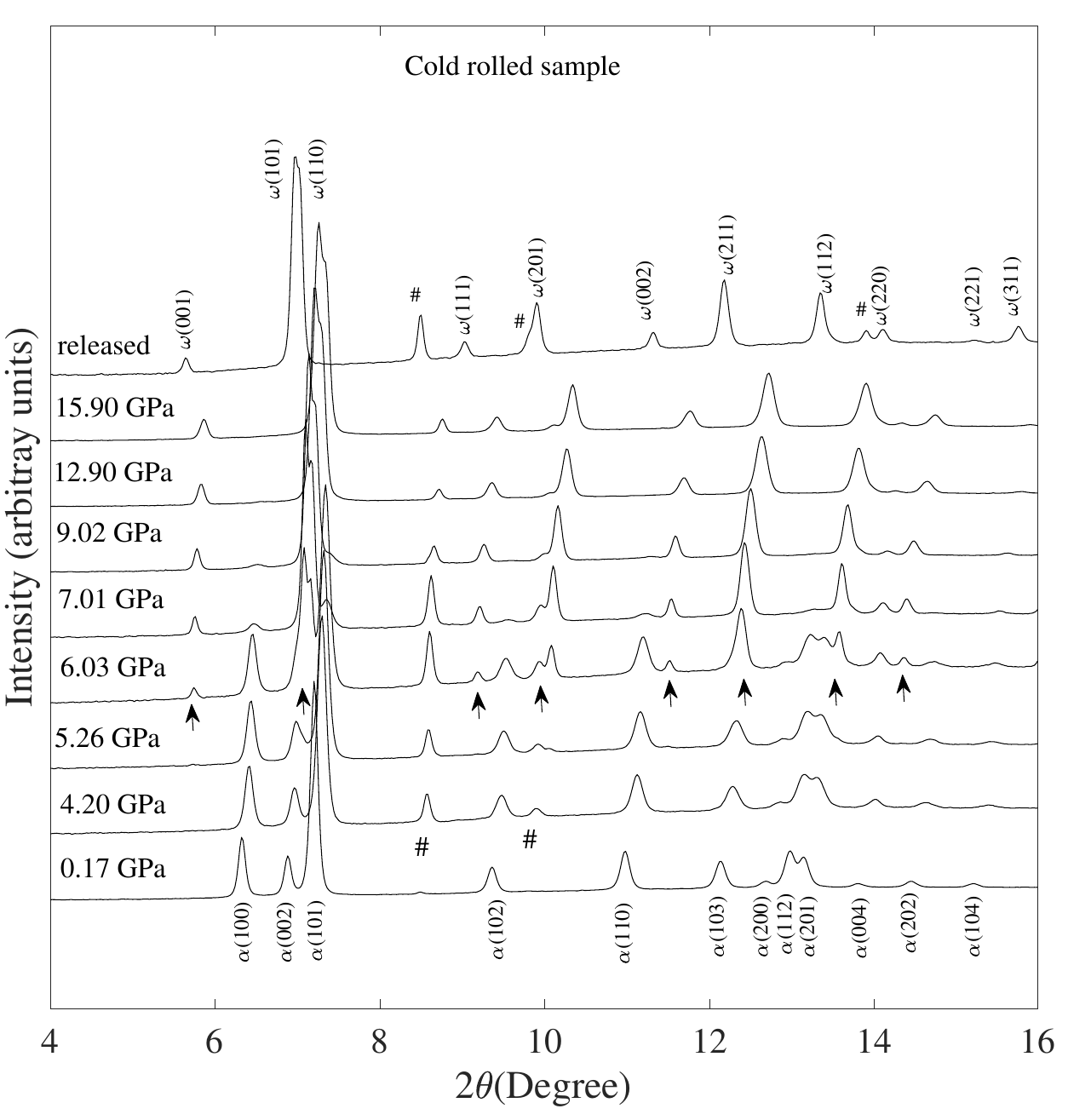}
\caption{High-pressure X-ray diffraction patterns of cold rolled sample at a few representative pressures. $\#$ mark indicates the (111) and (200) diffraction peaks of Cu. Arrow marks indicate diffraction peaks of $\omega$ Zr.  }
\label{fig:HPXRDstackcoldrolled}
\end{figure}

\begin{figure}[h!]
\includegraphics[width=0.8\linewidth]{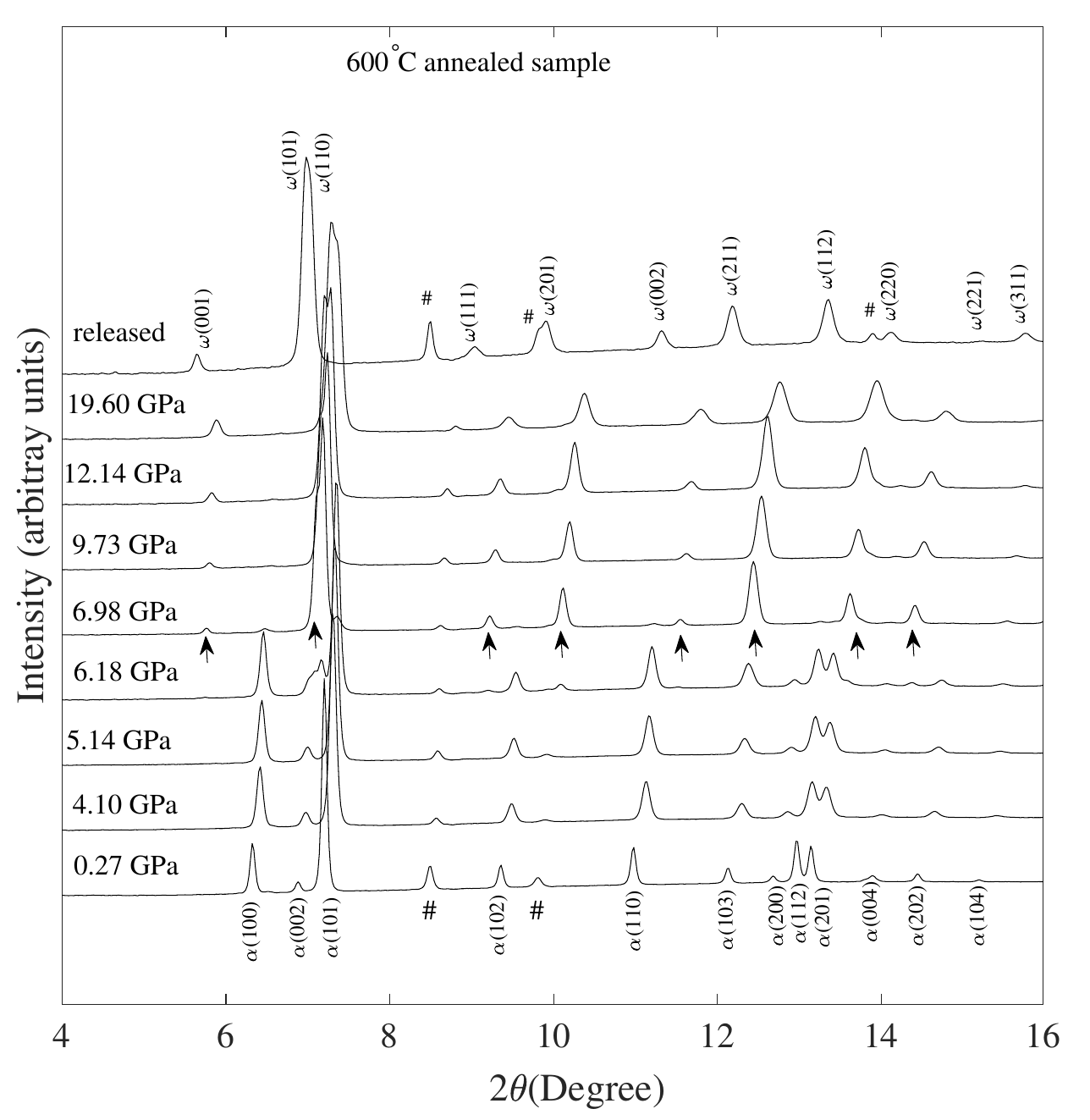}
\caption{High-pressure X-ray diffraction patterns of $600^\circ C$ annealed sample at a few representative pressures. $\#$ mark indicates the (111) and (200) diffraction peaks of Cu. Arrow marks indicate diffraction peaks of $\omega$ Zr.  }
\label{fig:HPXRDstackannealed}
\end{figure}

Figs. \ref{fig:HPXRDstackcoldrolled} and  \ref{fig:HPXRDstackannealed} show stacked diffraction patterns of cold rolled and $600^\circ C$ annealed samples at a few representative pressures. Both samples exhibit $\alpha\rightarrow\omega$ phase transition. In both cases, the $\omega$ phase does not transform back to the $\alpha$ phase on the complete release of pressure.

\subsection{Equations of state}
\begin{figure}[h!]
\includegraphics[width=1\textwidth]{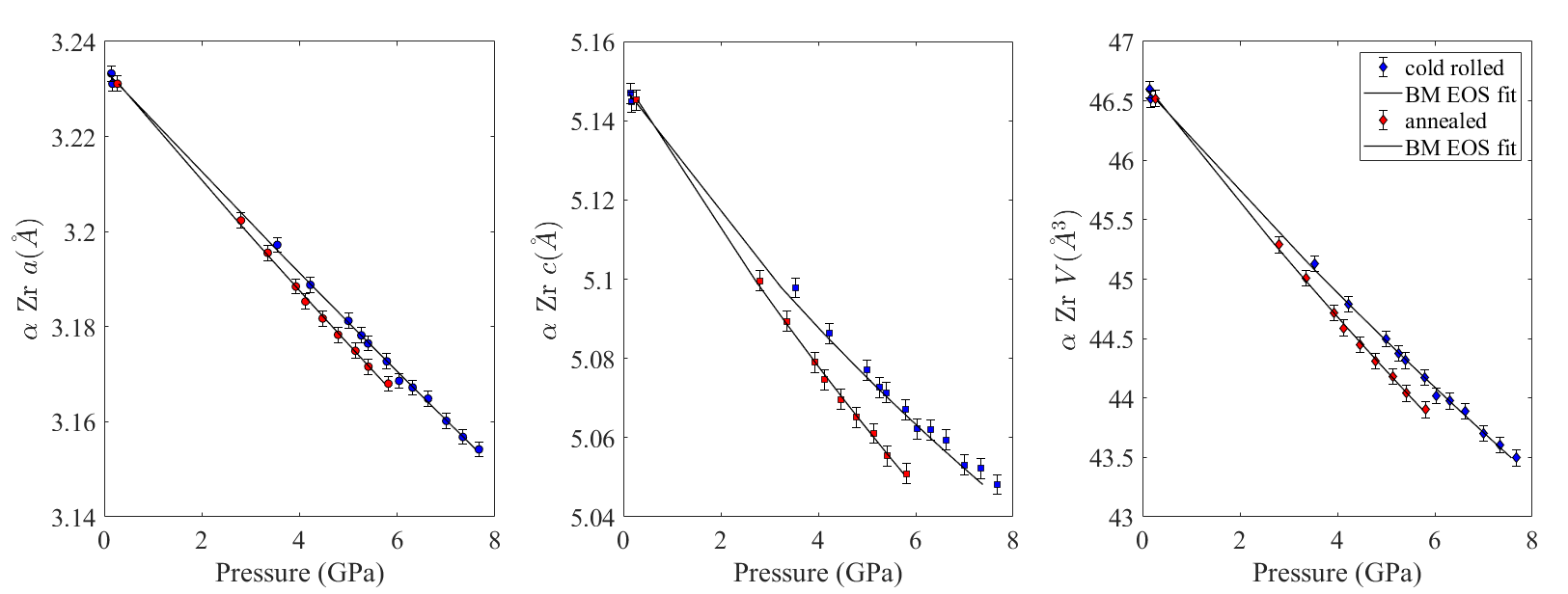}
\caption{Lattice parameters and unit cell volume as a function of pressure for $\alpha Zr$.}
\label{fig:pvplotalpha}
\end{figure}

\begin{figure}[h!]
\includegraphics[width=1\textwidth]{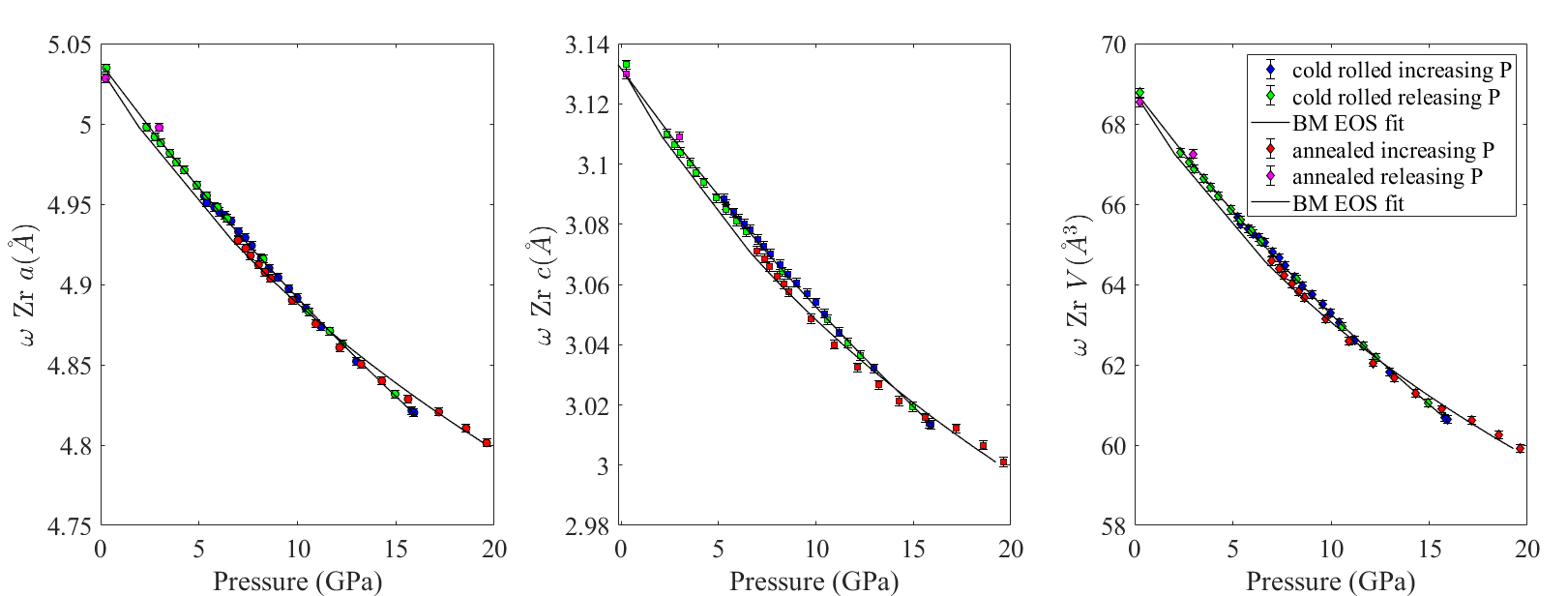}
\caption{Lattice parameters and unit cell volume as a function of pressure for $\omega Zr$.}
\label{fig:pvplotomega}
\end{figure}

These diffraction patterns were analyzed to obtain pressure-dependent lattice parameters, phase fractions, and microstructural evolution in each phase as a function of pressure. As can be seen in Figs. \ref{fig:pvplotalpha} the $p-a$, $p-c$, and $p-V$ data for pressure-dependence of the lattice parameters $a$ and $c$ and the unit cell volume $V$ for cold-rolled sample shows relatively lower compressibility as compared to the annealed sample for the $\alpha$ Zr phase. For $\omega$ Zr phase, the $p-a$, $p-c$, and $p-V$ data (Fig.  \ref{fig:pvplotomega}), are very close to each other for the cold-rolled and annealed samples within the experimental errors. The lower linear and bulk compressibility of the cold-rolled $\alpha$ Zr could be due to much smaller grain size (see below), more grain boundaries, and lower compressibility of the grain boundaries in comparison to the bulk crystal. 
For $\omega$ Zr, due to the irreversibility of transformation, we were able to determine all three curves during unloading down to zero pressure, i.e., including a metastability region. During loading and unloading, the behavior of $p-a$, $p-c$, and $p-V$ is very close but still with some scatter.


The 3$^{rd}$-order Birch-Murnaghan equation of state (BM-EOS) \cite{birch-1947} has been fitted to the $p-a$, $p-c$ and $p-V$ data for both $\alpha$ and $\omega$ phases of Zr to obtain bulk and linear moduli and their pressure dependence. The obtained results are given in Table \ref{BMfitresults}. {\color{black} Note that tensor $\fg F_*=\{a/a_0; a/a_0; c/c_0 \}$ characterizes an anisotropic deformation gradient state under hydrostatic loading for hexagonal and tetragonal crystals \cite{Levitas-PRB-21}.  It is utilized in \cite{Levitas-etal-NatCom-23} for finding analytical solutions and post-processing of the heterogeneous XRD patterns in the Zr sample plastically compressed without pressure transmitting medium and determination of the fields of stress components in the entire sample. It is now clear that data for tensor $\fg F_*$ for the cold-rolled Zr should be used in \cite{Levitas-etal-NatCom-23}.} While the  $p-a$, $p-c$, and $p-V$ behaviors for $\omega$  Zr do not significantly differ visually for the cold-rolled and annealed samples, it turns out the difference in parameters in the Birch-Murnaghan equation is essential. {\color{black} This is not important in the pressure range under study but may lead to essential differences while extrapolating to a higher pressure. To illustrate the quality of fit, the  $F-f$ plot for both the $\alpha$ and the $\omega$ phases for $600^\circ C$ annealed and cold rolled samples are given in the appendix \ref{BMeosfitting}.
 Scatter in data in Figs. \ref{fig:F-falpha}, \ref{fig:F-fomega},  \ref{fig:Ko-kpalpha}, and \ref{fig:Ko-kpomega} is, in particular, due to evolving microstructure and contributions of the defects in producing non-hydrostatic stresses, which affect EOS.  The larger scatter for the $\omega$ phase is because of very different microstructural parameters during pressure increase and decrease.
}


%
%

\begin{table}
\begin{center}

\caption{\label{BMfitresults}Bulk modulus, linear modulus, and their pressure derivatives obtained for $\alpha$ and $\omega$ Zr phases by fitting Birch-Murnaghan equation of state to $p-V$, $p-a$ and $p-c$ data using EOSFIT7-GUI software \cite{Rangel-2016}. The estimated standard deviation in each parameter is given in brackets.}
\begin{tabular}{c c c}
 \hline
 \hline
 Fitted parameters & cold rolled sample & $600^\circ C$ annealed sample \\
 \hline
 $\alpha$ Zr & & \\
 \hline
 Ambient unit cell volume $V_\circ$ ($\AA^3$) & 46.652(0.055) & 46.669(0.021)   \\
 Bulk modulus $K_o$ (GPa) & 100.26(5.57) & 86.90(3.21)\\
 Pressure derivative $K_p$ & 2.03(1.21) & 2.63(1.11)\\
 $a_\circ$ ($\AA$)& 3.2344(0.0012) & 3.2345(0.0005)\\
 Linear modulus $M_\circ$ (GPa) & 295.90(13.63)&265.51(9.26)\\
 Pressure derivative $M_\circ'$ & 2.06(2.70)&4.43(3.08)\\
 $c_\circ$ ($\AA$)& 5.1496(0.0029) & 5.1510(0.0009)\\
 Linear modulus $M_\circ$ (GPa) & 300.81(30.69)&248.22(11.45)\\
 Pressure derivative $M_\circ'$ & 21.93(8.74)&16.94(4.69)\\
 \hline
 $\omega$ Zr & & \\
 \hline
 Ambient unit cell volume $V_\circ$ ($\AA^3$) & 68.873(0.032) & 68.711(0.022)\\
 Bulk modulus $K_o$ (GPa) & 102.61(1.64) & 101.577(2.02)\\
 Pressure derivative $K_p$ & 3.19(0.22) & 3.29(0.39)\\
 $a_\circ$ ($\AA$)& 5.0378(0.0007) & 5.0328(0.0007)\\
 Linear modulus $M_\circ$ (GPa) & 303.57(4.31)&299.83(6.11)\\
 Pressure derivative $M_\circ'$ & 8.07(0.55)&9.32(1.14)\\
 $c_\circ$ ($\AA$)& 3.1336(0.0008) & 3.1324(0.0004)\\
 Linear modulus $M_\circ$ (GPa) & 315.23(9.36)&314.10(7.63)\\
 Pressure derivative $M_\circ'$ & 13.61(1.41)&11.25(1.54)\\
  \hline
\end{tabular}

\end{center}
\end{table}


\subsection{Phase transformation}
\begin{figure}[h!]
\includegraphics[width=0.6\linewidth]{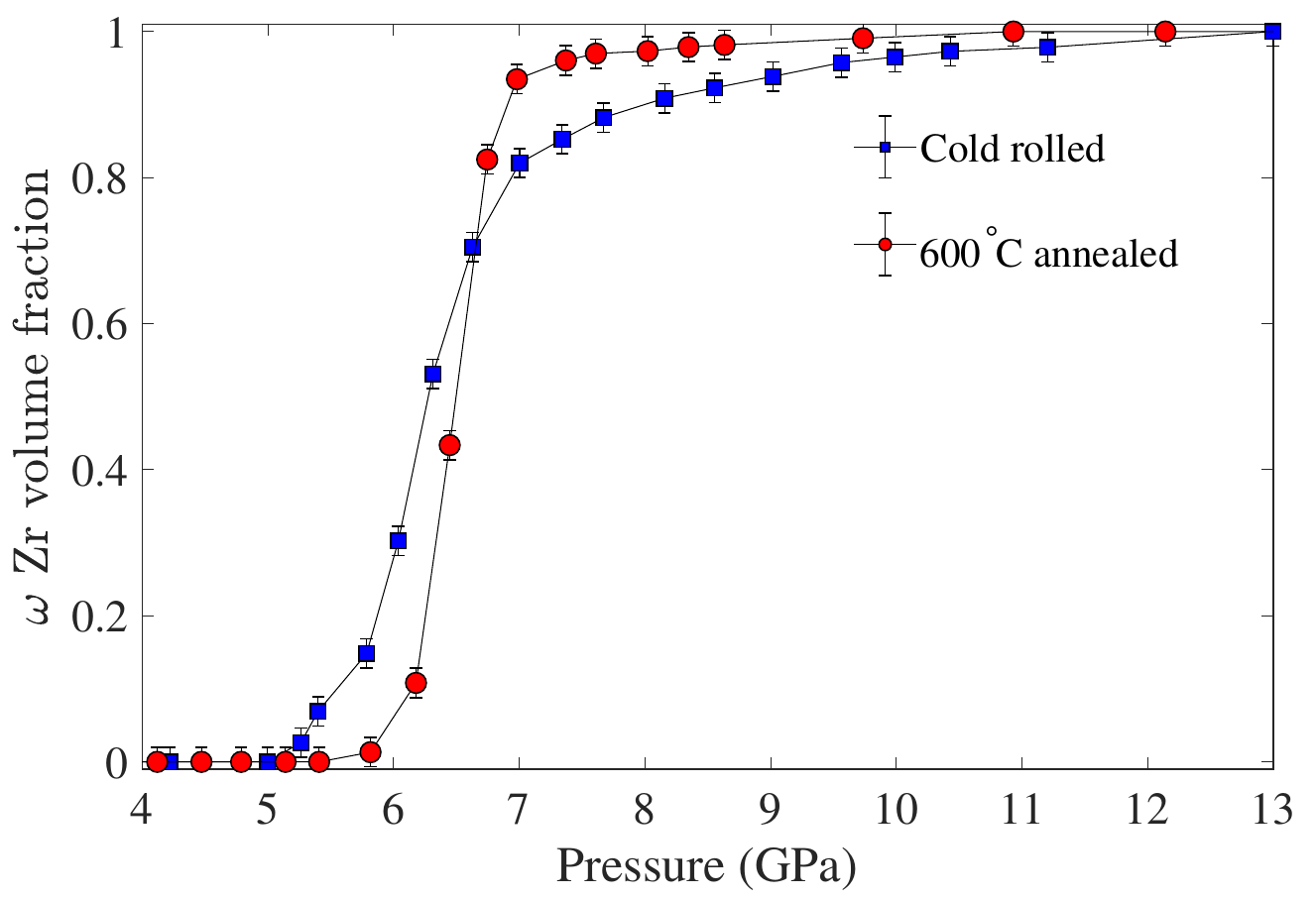}
\caption{Volume fraction of $\omega \, Zr$ phase as a function of pressure.}
\label{fig:pvscon}
\end{figure}
For cold-rolled sample with saturated hardness the $\alpha\rightarrow\omega$ phase transformation initiated at $\sim 5.1$ GPa and completed at $\sim 13.0$ GPa, whereas for the  $600^\circ C$ annealed sample, phase transformation initiation and completion pressures were  $\sim 5.9$ GPa and $\sim 10.9$ GPa respectively (Fig. \ref{fig:pvscon}). {\color{black} The intersection point of curves in Fig. \ref{fig:pvscon} is at a pressure of $\sim$ 6.6 GPa and volume fraction of $\omega$ Zr of 0.7} The results suggest that large plastic straining prior to transformation promotes {\color{black} multiple nucleations because of more and stronger stress concentrators (various dislocation configurations, twins, and grain boundaries)  but suppresses growth (i.e., reduces slope of the kinetic curve) by producing more obstacles (dislocation forest, point defects, grain boundaries) for interface propagation.} Reverse phase transformation is not observed down to zero pressure for both samples, therefore the hypothesis of symmetric hysteresis enlargement due to plastic straining cannot be tested here. However, the reduction in pressure for initiating the direct transformation contradicts the proportionality of the pressure hysteresis and hardness suggested in \cite{Blank-Estrin-2014} for multiple materials.
{\color{black} Consequently, general theory based on the equations $k=L \sg_y$   utilized in the phase transformation condition $X=k$,
and analytical and numerical solutions in \cite{Levitas-JMPS-I-97,Levitas-JMPS-II-97,levitas-ijss1998,Levitas-IJP-21,Levitasetal-PhilMag-02}  must be essentially advanced.
}
A possible reason for this discrepancy is that results in \cite{Blank-Estrin-2014} (see also more experimental detail in \cite{Blanketal-InorMat-83}) have been obtained in piston-cylinder by recording force-displacement curves
without in situ XRD probing. Pressure is determined by force using calibration based on known pressure for phase transformation in bismuth. Thus, nucleation of a small amount of high-pressure phase could not be detected and the determination of pressure has significant error.  Our results on reduction in pressure for initiation of $\omega$ Zr in commercially pure Zr
are consistent with those for extra pure Zr in \cite{Kumaretal-Acta-20}
after plastic compression by $5\%$ and $10\%$ prior to transformation.
However, in \cite{Kumaretal-Acta-20}
 the entire kinetic curve $c(p)$ is shifted to lower pressures with increasing plastic strain, i.e., there is no suppression of growth. This shows that the effect of plastic strain is non-monotonous, i.e., the small plastic strain does not produce obstacles for growth, but the large plastic strain, significantly reducing grain size and increasing the dislocation density, suppresses it.
  {\color{black} Also, reduction in PT pressure in \cite{Kumaretal-Acta-20} is larger for straining by $5\%$ and then reduces for straining by $10\%$, which is difficult to explain from the physical point of view. However,  the experiments in  \cite{Kumaretal-Acta-20} were performed in multi-anvil apparatuses without hydrostatic pressure-transmitting medium, i.e., in the presence of non-controllable non-hydrostatic stresses.  This may be the reason for the above result. Also, in situ evolution of the microstructure was not recorded in  \cite{Kumaretal-Acta-20} and pressure was estimated based on XRD measurements in an alumina piston, which is less precise than based on a Cu marker placed in the same pressure transmitting medium like a sample, as we did here. }
  In any case, more detailed microstructure-related studies are required.

\subsection{Microstructure evolution}

\begin{figure}[h!]
\includegraphics[width=0.6\linewidth]{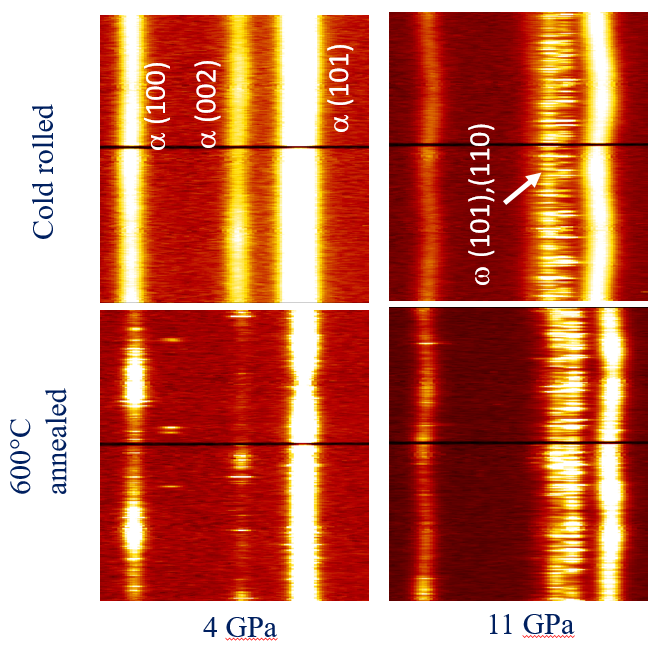}
\caption{Cake view of X-ray diffraction images recorded at 4 GPa and 11 GPa for the cold-rolled and $600^\circ C$ annealed samples.}
\label{fig:imagecake}
\end{figure}

A closer look at the diffraction images (Fig. \ref{fig:imagecake}) shows that the diffraction rings are relatively smooth and broad for the cold-rolled sample as compared to the annealed sample, which suggests finer domain size in the cold-rolled sample. For the annealed Zr the diffraction ring becomes smoother, suggesting a reduction in domain size with pressure. Spotty data of Zr-$\omega$ phase is suggestive of grain growth across phase transition (see below), in agreement with previous experiments
\cite{Velisavljevic-etal-11,Popovetal-SciRep-19}.

\begin{figure}[h!]
\includegraphics[width=\linewidth]{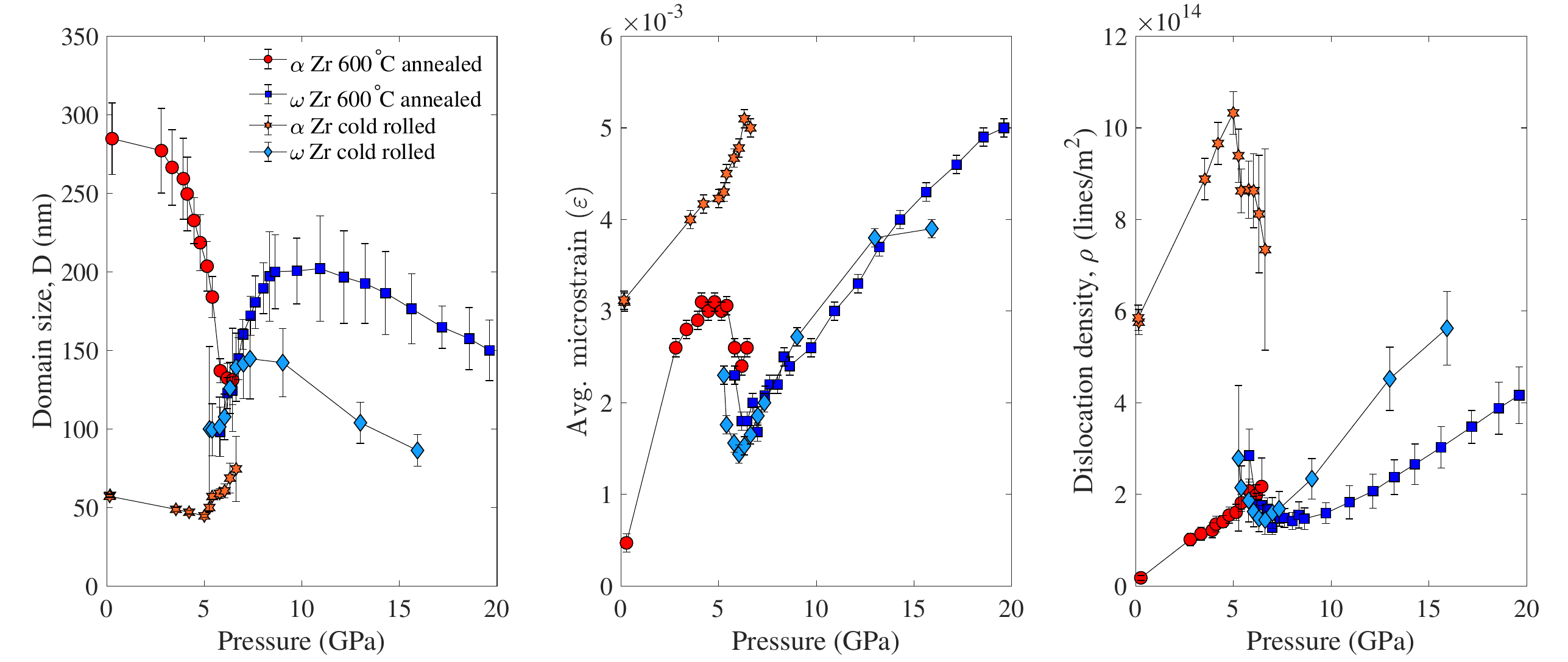}
\caption{Average domain size, microstrain, and dislocation density in each phase of Zr for the cold rolled and $600^\circ C$ annealed samples as a function of high pressures.}
\label{fig:MSresults}
\end{figure}

{\color{black} Fig. \ref{fig:MSresults}  shows the results from microstructural analysis of XRD data versus pressure for both $\alpha$ and $\omega$ Zr before, during, and after the phase transformation.  They will be analyzed in Sections \ref{alphaZr}, \ref{alpha-omegaZr}, and \ref{omegaZr}.  Fig. \ref{fig:MS-c-results} shows the same parameters in both phases across
 the phase transformation versus pressure, which will be discussed in Sections \ref{MA-vs-c}.

 Note that twinning is essentially involved in the plastic deformation of Zr  \cite{Kumaretal-Acta-20,Tome-2009,Brown-2006,Tome-2001,Pollock-2013}. Since twins can be presented as arrays of partial (twinning) dislocations at their interfaces, they are caught by the XRD approach as well.}


{\color{black}
\subsubsection{$\alpha$ Zr before phase transition\label{alphaZr}}
For the cold-rolled sample, domain size  decreases from $\sim 60\, nm$ to $\sim 45\, nm$ before phase transition whereas substantial reduction in domain size $D$ has been observed for the annealed sample from $\sim 300\, nm$ to $\sim 120\, nm$,
with a very sharp reduction between 3.5 and 5.9 GPa.
Dislocation  density $\rho_d$  in $\alpha$ phase increases linearly with pressure from $6$ to $11 \times 10^{14}\, lines/m^2$
for the cold-rolled sample and from $0.2$ to $2.1 \times 10^{14}\, lines/m^2$
for the annealed sample.
Note that results for the cold-rolled sample are comparable to those in \cite{Lin-Levitas-MRL-23}. Thus,
after multiple rolling, with $\rho_d =10^{15}  lines/m^2$ and $D = 75\, nm$;
after plastic compression with smooth anvils, immediately before
initiation of the strain-induced $\alpha-\omega$ phase transformation at 1.36
GPa, with $\rho_d =1.26  \times 10^{15}  lines/m^2$ and $D = 65\, nm$; after compression with rough diamond anvils, which intensify plastic flow, also just before initiation
of the strain-induced $\alpha-\omega$ transformation at 0.67 GPa, $\rho_d =1.83 \times 10^{15}  lines/m^2$ and $D = 48\, nm$.
This finding has two important consequences for nanostructured Zr produced by severe plastic deformation at normal pressure, e.g., by rolling. (a)
High hydrostatic pressure produces further domain refinement and an increase in dislocation density similar to large plastic deformations with increasing pressure. (b) Hydrostatic pressure allows to easily further refine grain and increase dislocation density in  $\alpha$ Zr,
while plastic straining causes transformation to $\omega$ Zr at low pressure.

The simplest Orowan's equation for plastic shear is $\gamma_p= \rho_d b l$, where $b $ is the modulus of the Burger's vector, and $l$ is the averaged
travel distance of dislocation \cite{Meyers-Chawla-02}. An increase in dislocation density (assuming for simplicity that a new dislocation loop travels by the same distance $l$ as already existing dislocations while forming) produces an increment of plastic deformation. If domain boundaries are small-angle boundaries, they can be represented as dislocation walls.  When domain size reduces, the dislocation wall moves covering distance $l$, producing additional plastic strains.
For a rough estimate of plastic shear, let us use for the cold-rolled sample increase in dislocation density of $5 \times 10^{14}\, lines/m^2$
with $l=50\, nm$ (intermediate domain size). Substituting these numbers along with $b=0.32\, nm$ (using intermediate lattice parameter) in Orowan's equation, we obtain
$\gamma_p=0.80 \%$. Similarly, for the annealed sample, an increase in dislocation density of $1.9 \times 10^{14}\, lines/m^2$ with $l=200\, nm$ result in $\gamma_p=1.21 \%$. In reality, dislocations may annihilate and be absorbed by grain boundaries, and new dislocations are generated, moved, and absorbed again,
thus they pass domain size multiple times, and the resultant plastic shear could be much larger.
Thus,
hydrostatic pressure produces essential plastic strain, which can be characterized in terms of change in   $\rho_d$ and also domain size.
Since there is no preferred direction under hydrostatic loading,    these heterogeneous shears compensate each other and are geometrically invisible at the scale of the sample. A similar situation is for martensitic phase transformation under hydrostatic loading or cooling/heating. While each crystal has significant transformation shear, they compensate each other and are geometrically invisible at the scale of the polycrystalline sample.
The physical reason for plastic deformation under hydrostatic loading is
that various defects (dislocations, twins, grain boundaries, and their junctions) cause significant local internal stresses with deviatoric/shear stress components reaching the yield strength, which causes plastic deformations.
In addition, dislocation core energy and structure, and elastic moduli change with pressure, which may lead to redistribution of dislocation configurations, increase in dislocation density, domain and grain refinement, etc.

The microstrain  in $\alpha$ phase increases from $0.0031$ to $0.0042$  at the transition point with some saturation near the transition point for the cold-rolled sample, and from $0.0005$ to $0.0032$ with some saturation and then drops to
$0.0026$ at the transition point for the annealed sample. We will use these data while rationalizing the phase transformation pressures
for both sample types.

\subsubsection{Microstructure evolution in $\alpha$ and $\omega$ Zr during phase transition versus pressure \label{alpha-omegaZr}}
The domain size for $\omega$ phase at the initiation of phase transformation is $\sim 100\, nm$ for both cases and increases to a maximum $\sim 140\, nm$ and $\sim 200\, nm$ for cold-rolled and annealed samples, respectively at $\sim$ 8 GPa pressure (Fig. \ref{fig:MSresults}).
Surprisingly, the initial domain size of $\omega$ phase, $\sim 100\, nm$, and the growing parts of the curves up to $\sim 145\, nm$ are independent of the plastic strain tensor prior to transformation. The initial domain size  of $\sim 100\, nm$  probably characterizes
size of operational nucleus, which will grow rather than collapse \cite{olson+cohen-1986}; it is smaller than $\sim 120\, nm$ in $\alpha$ phase at the initiation of the transformation for the annealed material and larger than $\sim 45\, nm$ for the cold-rolled material.
Observed growth of the domain size in $\omega$ Zr during phase transformation
is presumably related to the growth of the transformed regions.
After reaching the maximum in $D$  for both cases, there is a plateau followed by monotonous domain size reduction up to the completion of the transformation.

The domain size of $\alpha$ Zr at the initial stage of transformation reduces for the annealed sample and increases for the cold-rolled sample.
This may indicate the different nucleation mechanisms for annealed and cold-rolled samples. There is always quite a broad distribution
of the domain sizes.  Reduction of the domain size of $\alpha$ Zr for annealed sample
may be caused by nucleation mostly in large grains, reducing the size of the averaged $\alpha$ Zr domains. This corresponds to the
general trend that under hydrostatic loading, the transformation pressure for various materials increases with the reduction of the grain size.
  In contrast, an increase in
 the domain size in $\alpha$ Zr for the cold-rolled sample
may be caused by transformation mostly in small grains, eliminating them from the contribution to the averaged size.
This transformation is caused by additional stress concentrators generated by severe plastic deformation.

The dislocation density in the $\omega$ phase at the beginning of the transformation is much lower than in  $\alpha$ Zr for the cold-rolled sample and
larger than in  $\alpha$ Zr for the annealed sample. That means that dislocation structure is not inherited during this reconstructive transformation.  This may happen if moving $\alpha$-$\omega$ interfaces sweep away the entire microstructure (domains and dislocations) in the $\alpha$ phase and new domains and dislocations are formed in the $\omega$ Zr.
 In the $\omega$ phase, during the initial stage of phase transformation, dislocation  density  reduces
from $2.78$ to $1.4 \times 10^{14}\, lines/m^2$
for the cold-rolled sample and from $2.84$ to $1.39 \times 10^{14}\, lines/m^2$
for the annealed sample. That means that it is independent of the  severe plastic strain tensor prior to transformation, similar to the
domain size.
Then there is a plateau for the annealed sample followed by growth, much slower than for the cold-rolled, with
$1.39 \times 10^{14}\, lines/m^2$ at the end of transformation for  the annealed sample versus $2.74 \times 10^{14}\, lines/m^2$
for the cold-rolled sample. This is consistent with the variation of the domain size during transformation.

 During transformation, dislocation  density   in $\alpha$ phase reduces with pressure down to $7.5 \times 10^{14}\, lines/m^2$ for the cold-rolled sample and practically does not change
for the annealed sample.

The microstrain in the $\omega$ phase is nearly the same at the transformation initiation pressure for both cases, $0.0023$ for pre-deformed and $0.0022$  for annealed samples.
It  reduces during phase transformation  to $0.0013$ for the cold-rolled sample and $0.0017$ for the annealed sample,
demonstrating stress relaxation caused by transformation strain; it is not transformation-induced plasticity, because dislocation density reduces at this stage.
Then it monotonously increases for both cases. Surprisingly,  the {\it microstrain versus pressure is practically independent of the plastic strain tensor prior to transformation during the entire transformation.} More general rule will be formulated below in terms of volume fraction of
$\omega$ phase instead of pressure.
In the $\alpha$ phase during the phase transformation, microstrain remains the same for the annealed sample and grows in the pre-strained sample, showing that due to smaller yield strength in the annealed sample,
stress relaxation is much more pronounced than in the pre-strained sample.

\subsubsection{Microstructure evolution across the phase transformation versus volume fraction of $\omega$ Zr\label{MA-vs-c}}
\begin{figure}[h!]
\includegraphics[width=\linewidth]{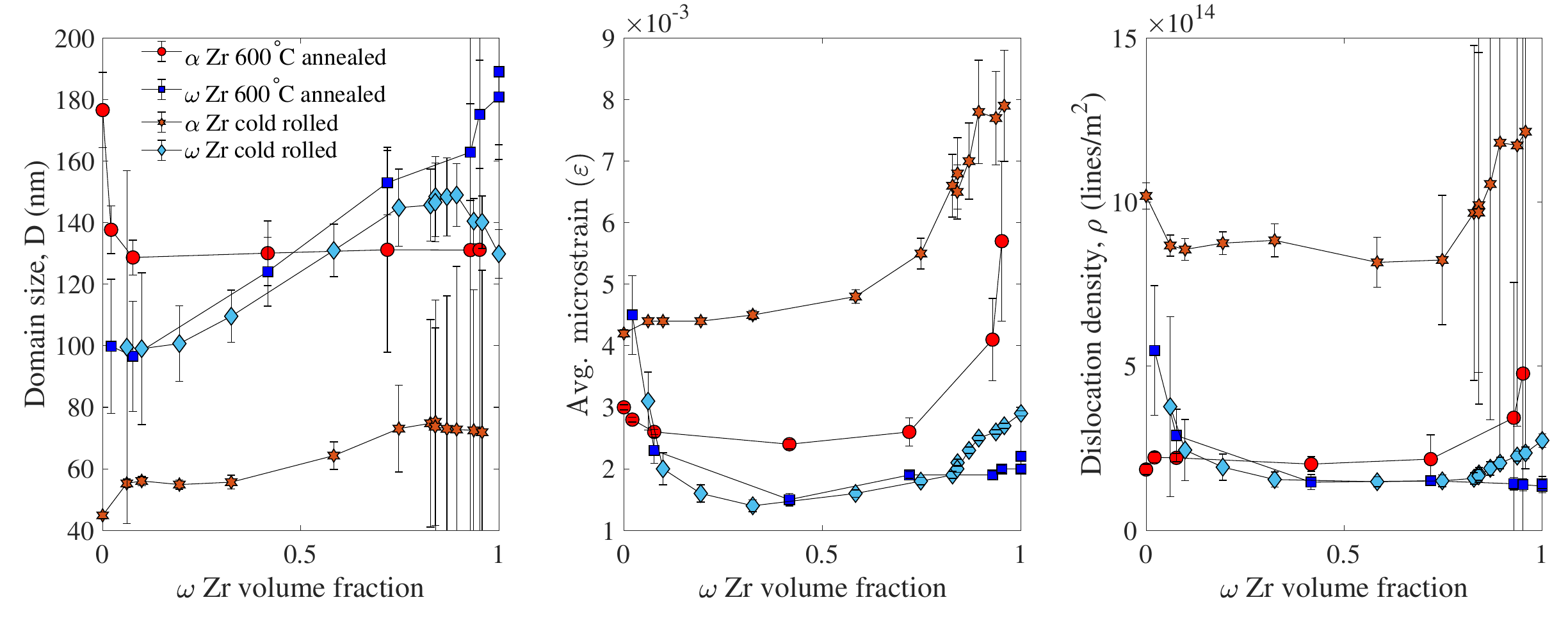}
\caption{Average domain size, microstrain, and dislocation density in $\alpha$ and $\omega$ Zr for the cold rolled and $600^\circ C$ annealed samples as a function of volume fraction of $\omega$ Zr.}
\label{fig:MS-c-results}
\end{figure}

The evolution of all three microstructural parameters in terms of the volume fraction of $\omega$ Zr (Fig. \ref{fig:MS-c-results}) looks much more informative than versus
pressure. In particular,  for all three microstructural parameters in $\omega$ Zr results for the cold rolled and annealed samples are very close for $c<0.8$.
This constitutes the following rule:
{\it The average domain size, microstrain, and dislocation density in  $\omega$ Zr for $c<0.8$ are functions of the volume fraction of $\omega$ Zr only, which
are independent of the plastic strain tensor prior to transformation and pressure. }

  For the $\alpha$ Zr, the following results have been obtained.

The domain size of $\alpha$ Zr at the initial stage of transformation reduces for the annealed sample and increases for the cold-rolled sample.
Then the domain size  in $\alpha$ Zr  remains constant through the entire transformation for the annealed Zr and
has plateaus for $0.06<c< 0.35$  and $0.8<c< 1$ with the growing portion between.
Dislocation density for the annealed remains constant up to $c=0.71$, then increases. For the cold-rolled Zr, it first reduces, then remains approximately constant, and sharply increases toward the end of the transformation. Microstrain has roughly similar trends for both samples, with
larger values for the pre-strained sample.

\subsubsection{ $\omega$ Zr after completing  phase transition and during pressure release\label{omegaZr}}
In the single $\omega$ phase region, domain size reduces with pressure, continuing the curves during transformation.
Dislocation density  increases to $5.6 \times 10^{14}\, lines/m^2$ for the cold-rolled sample at 15.9 GPa (the last available point for the pre-deformed sample) and to $3.03 \times 10^{14}\, lines/m^2$ at 15.6 GPa and
$4.2 \times 10^{14}\, lines/m^2$ at 19.6 GPa for annealed sample.
The microstrain increases to $0.0039$ for both samples at 15.9 GPa (the last available point for the pre-deformed sample) and continues
growing to $0.0050$ for the annealed sample at 19.6 GPa.
}

{\color{black}
Microstructure and properties after pressure release are of paramount importance for practical applications.
The microstructural evolution for the annealed and the cold-rolled samples with increasing and decreasing pressure, shown in Figs. \ref{fig:MS-annealed-inc-dec} and \ref{fig:MS-cold rolled-inc-dec}, contains nontrivial results.
On pressure release, average domain size remains nearly the same as at the highest measured pressure, i.e., $\sim 150\, nm$ and $\sim 80\, nm$ for the annealed and the cold-rolled samples, respectively.
Such irreversibility in the domain size is of practical meaning for producing nanograined materials. While the final domain size for the annealed sample is reduced from $\sim 300\, nm$ in $\alpha$ phase to $\sim 150\, nm$ in $\omega$ phase due to high-pressure treatment and phase transformation, for plastically pre-strained sample it increased from $\sim 60\, nm$ in $\alpha$ phase to $\sim 80\, nm$ in $\omega$ phase.

The average microstrain and the dislocation density for the annealed sample,  while remaining unchanged down to 3.5 GPa, reduce from $5.0$ to $4.2\times 10^3 \,lines/m^2$ and from $4$  to $3.4 \times10^{14}\, lines/m^2$, respectively, after complete pressure release.
Both final values are an order of magnitude larger than for the initial $\alpha$ phase.
Similar behavior is observed for the dislocation density for the cold-rolled sample but with the final value in the $\omega$ phase 2 times smaller than in the initial $\alpha$ phase. More complex variation for the cold-rolled sample is found for
the average microstrain: it slightly reduces when the pressure drops from 15 to 11 GPa, then remains constant till 3.5 GPa, and reduces from $3.4$ to $2.1\times 10^3$ with full pressure release, which is 1.5 times smaller than for the initial $\alpha$ phase. Surprisingly,
for the annealed sample, the final dislocation density and average microstrain (and the domain size, which is, however, expected) in the $\omega$ phase are larger than for the severely pre-deformed sample. Thus, if the goal is to improve the microstructure due to pressure treatment and irreversible phase transformation,
it should be done on the annealed sample for the dislocation density and average microstrain and on the severely pre-strained sample for the domain size. Based on the obtained rules, one can easily determine from Figs. \ref{fig:MS-annealed-inc-dec} and \ref{fig:MS-cold rolled-inc-dec}
the final microstructural parameters for unloading from lower pressures. Also, one can conclude that post-mortem measurements after pressure release give correct results for high pressures for domain sizes, slightly underestimate values of the dislocation density and average microstrain for the anneal sample, and strongly underestimate the dislocation density and microstrain for the cold-rolled sample.


\begin{figure}[h!]
\includegraphics[width=\linewidth]{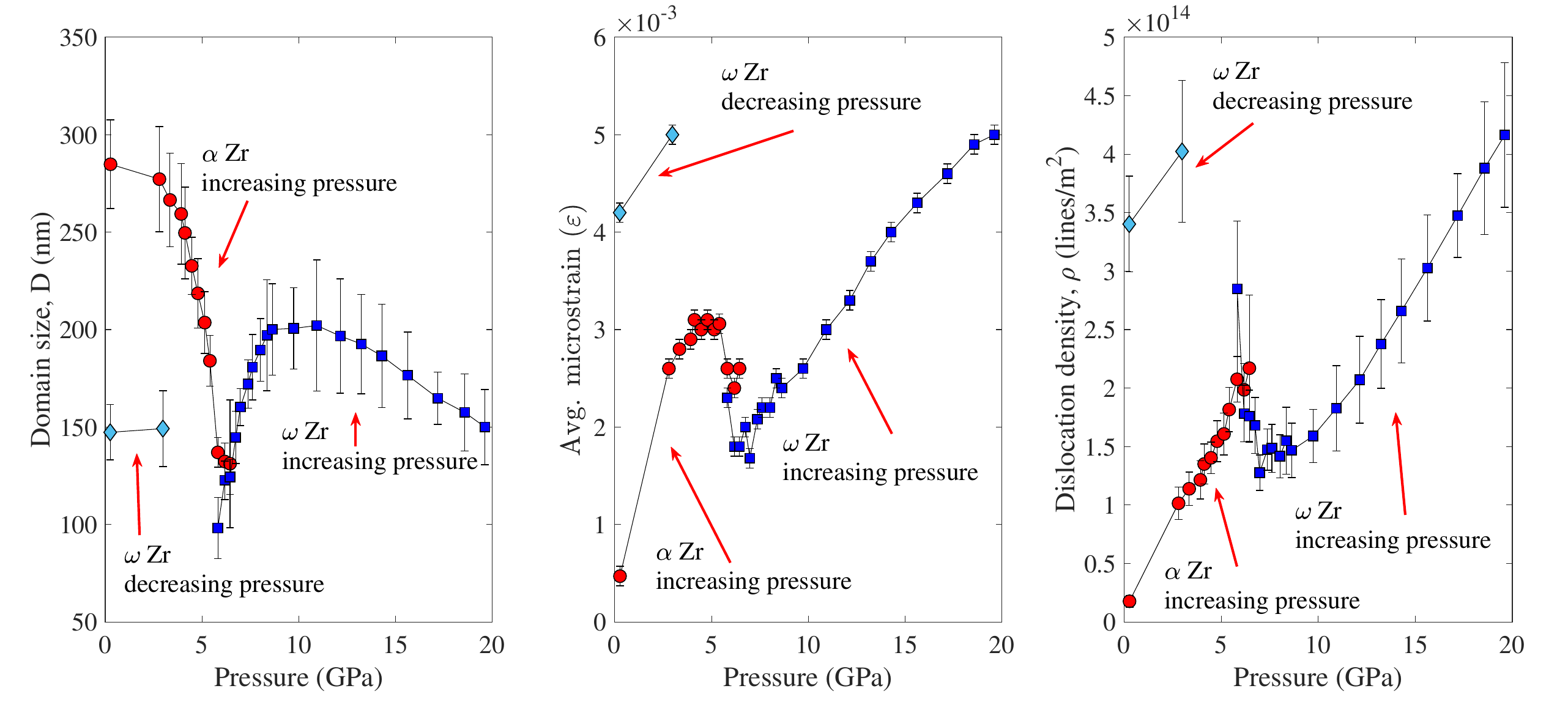}
\caption{Average domain size, microstrain, and dislocation density in $\alpha$ and $\omega$ Zr for the  $600^\circ C$ annealed sample while increasing and decreasing pressure.}
\label{fig:MS-annealed-inc-dec}
\end{figure}

\begin{figure}[h!]
\includegraphics[width=\linewidth]{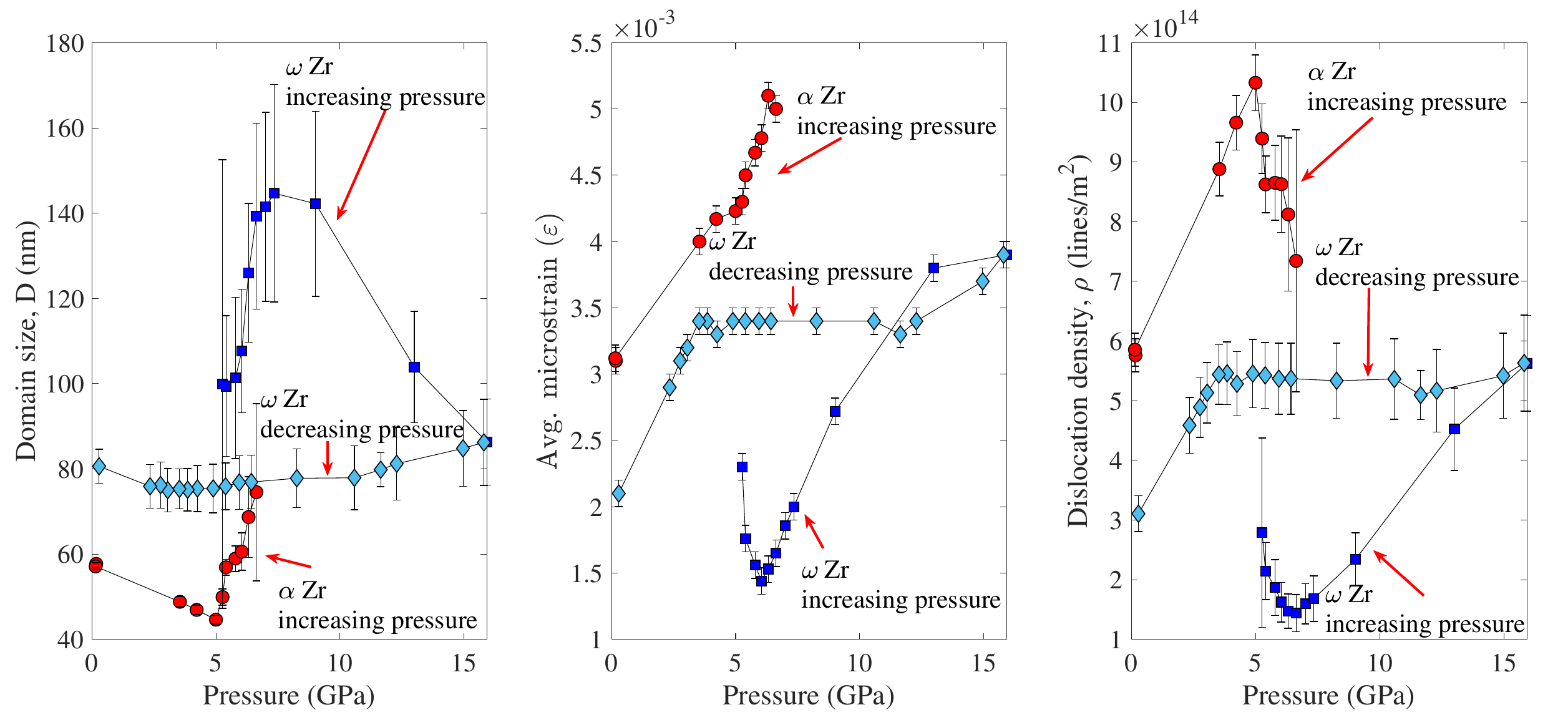}
\caption{Average domain size, microstrain, and dislocation density in $\alpha$ and $\omega$ Zr for the cold-rolled sample while increasing and decreasing pressure.}
\label{fig:MS-cold rolled-inc-dec}
\end{figure}
}

All the above results demonstrate the essential and complex microstructure evolution for single-phase samples and during phase transformation and pressure cycling.
The significant evolution of the microstructure and its effect on phase transformation demonstrates that their postmortem evaluation does not represent the actual conditions during loading.

{\color{black}
\subsection{Rationalizing the effect of the microstructure on the EOS and phase transformation}
We assume that among three microstructural parameters, domain size, microstrain, and dislocation density, the microstrain plays the dominant role in EOS. Indeed, EOS for $\omega$ Zr for cold-rolled and annealed samples are close (Fig.  \ref{fig:pvplotomega}), similar to the microstrain versus pressure (Fig. \ref{fig:MSresults}). At the same time,  the domain size and dislocation density versus pressure is essentially different for the cold-rolled and annealed samples (Fig. \ref{fig:MS-c-results}). Then the difference in EOS for $\alpha$ Zr for the cold-rolled and annealed samples can also be connected to different microstrains.  Microstrain characterizes internal strains and stresses with significant non-hydrostatic components. It is well known that the EOS under hydrostatic and non-hydrostatic loadings are different, which explains the difference in the EOS for
$\alpha$ Zr for the cold-rolled and annealed samples.  Scatter in data in Figs. \ref{fig:F-falpha}, \ref{fig:F-fomega},  \ref{fig:Ko-kpalpha}, and \ref{fig:Ko-kpomega} is, in particular, due to evolving microstructure and contributions of the defects in producing non-hydrostatic stresses, which affect EOS. The larger scatter for $\omega$ phase is because of very different microstructural parameters during pressure increase and decrease (Fig. \ref{fig:MS-annealed-inc-dec} and \ref{fig:MS-cold rolled-inc-dec}).
}

In the attempt to rationalize phase transformation initiation conditions using the above results,
let us assume that averaged microstrain in $\alpha$ phase $\varepsilon_l$ generates local volumetric
strain $3 z \varepsilon_l$ and maximum internal pressure at the phase transformation point of $3 z K(p) \varepsilon_l$, for both samples, where $z>1$ is the strain concentration factor.
Assuming that the total local pressure at the phase transformation point, $p_l$ in both samples is the same,
 \bey
p_{l1}=p_1+3 z K(p_1) \varepsilon_{l1} =p_{l2}=p_2+3 z K(p_2) \varepsilon_{l2}
\quad \rightarrow \quad
z= \frac{p_2-p_1}{3(K(p_1) \varepsilon_{l1}-K(p_2) \varepsilon_{l2})},
 \label{AM-1}
 \eey
where subscripts 1 and 2 designate cold-rolled and annealed samples, respectively, and $p_i$ are the applied pressure
for the initiation of the phase transformation for these samples. Substituting $p_1=5.1 $ GPa, $p_2=5.9 $ GPa, $\varepsilon_1=0.0042$,
 $\varepsilon_2=0.0026$, $K(p_1) =110.61  GPa$, $K(p_2) =102.39  GPa$, we obtain $z=1.34$, which is reasonable.

 Extension of the model based on the microstrain for the growth stage is impossible because microstrain in $\alpha$ phase increases for the pre-deformed sample and remains constant in the annealed sample, but growth in the pre-deformed sample is slower than in the annealed sample. The transformation pressure for the growth model can be based on the initial  hardness or yield strength
 $\sg_y$
 (like in \cite{Levitas-JMPS-I-97,Levitas-JMPS-II-97,levitas-ijss1998,Levitasetal-PhilMag-02,Levitas-JPCM-18}).
 Since  $\sg_y=\sg_{y0} + A \sqrt{1/D} + B \sqrt{\rho_d} $, where $D$ is the domain size and $\rho_d$ is the dislocation density, resistance to growth can be connected to the domain size and dislocation density.
 The second term in the expression for $\sg_y$ is due to the Hall-Patch effect and the third term is Taylor's strain hardening. The relationship between  the deviation of the actual phase transformation pressure from the phase equilibrium pressure
 during the growth stage and the yield strength
 seems reasonable because the deviation
characterizes the resistance to a moving interface
due to the material’s microstructure and the yield strength characterizes
the resistance to the motion of dislocations through
the same microstructure. More experiments are required for the calibration of the growth model.

\section{Concluding remarks}
{\color{black}
Experiments under hydrostatic conditions are considered as "clean" testing under well-defined pressure without deviatoric stresses and, consequently, plasticity. It is generally accepted in plasticity theory that
hydrostatic loading of void-free materials does not cause plastic deformation. Phase transformation pressures in literature have a very large scatter for the same material. In most cases, the initial microstructure and its evolution are not characterized, implying that they are not important for EOS and phase transformations.
In this paper, we address all these points and confront some well-known paradigms in plasticity theory and the theory of phase transformations
for two samples of commercially pure Zr: after cold-rolling to the saturated maximum hardness and annealing at  $600^\circ C$, respectively.

The crystal domain size significantly reduces and microstrain and dislocation density increase during loading for both  $\alpha$ and $\omega$ phases in their single-phase regions. For the $\alpha$ phase,
domain sizes are much smaller for prestrained Zr, while microstrain and dislocation densities
are much higher. Despite the generally accepted concept that hydrostatic pressure does not cause plastic straining,
a change in dislocation density and domain size implies that it does.
Simple estimate offers plastic shear values in $\alpha$ Zr as $0.75 \%$  for the prestrained Zr and $1.14 \%$ for the annealed Zr.
Since there is no preferred direction under
hydrostatic loading, these heterogeneous local shears are mutually compensated and are geometrically invisible at the scale of
the sample.
Consequently, classical plasticity theory should be advanced to include plasticity at hydrostatic conditions,
characterized in terms of microstructural parameters instead of geometric changes.
The physical reason for plastic deformation under hydrostatic loading is
that various defects (dislocations, twins, grain boundaries, and their junctions) cause significant local internal stresses with deviatoric/shear stress components reaching the yield strength, which causes plastic deformations.
In addition, dislocation core energy and structure, and elastic moduli change with pressure, which may lead to redistribution of dislocation configurations, increase in dislocation density, domain and grain refinement, etc.

For the maximum pressure of 5.9 GPa in a single-phase $\alpha$ Zr in the cold-rolled sample, the domain size
reduced to $\sim 45\, nm$ and dislocation density increased to $1.1 \times 10^{15}\, lines/m^2$. Such values are desirable for nanostructured materials and usually are reached by severe plastic deformation under high pressure. Consequently, hydrostatic compression of strongly pre-deformed Zr can be used for grain refinement, with some advantages that plastic straining may lead to a brittle  $\omega$ Zr at relatively low pressure, which is not always desirable.
}
PT $\alpha\rightarrow\omega$  initiates at a higher pressure for the annealed sample as compared to the cold-rolled sample,
in agreement with experiments in \cite{Kumaretal-Acta-20} for small plastic straining prior to transformation and in contrast to
general regularity suggested in \cite{Blank-Estrin-2014} for multiple materials and used in the models
in \cite{Levitas-JMPS-I-97,Levitas-JMPS-II-97,levitas-ijss1998,Levitasetal-PhilMag-02,Levitas-JPCM-18}.   This implies that  plastic straining prior to transformation promotes nucleation with more and stronger stress concentrators (various dislocation configurations, twins, etc), in agreement with analytical \cite{olson+cohen-1986,levitas-mechchem-04,levitas-prb-04}
 and computational    \cite{Xu-Khachaturyan-etal-NPJ-CM-18,Xu-Khachaturyan-etal-AM-19,levitas-javanbakht-APL-13,Javanbakht-Levitas-JMS-18}
studies.
With phase transformation progress, the promoting effect of prior straining reduces with crossover to suppressing effect at $c=0.7$ (pressure $\sim$6.6 GPa), above which volume fraction of $\omega$ Zr for the same pressure is higher for the annealed sample.
  Completion pressure for $\alpha\rightarrow\omega$  transformation is higher for the cold-rolled sample than for the annealed sample by 2 GPa suggesting that prior straining suppresses growth by producing more obstacles (dislocation forest, point defects, domain, and grain boundaries) for interface propagation.  This implies that the general theory \cite{Levitas-JMPS-I-97,Levitas-JMPS-II-97,levitas-ijss1998,Levitasetal-PhilMag-02,Levitas-JPCM-18} based on the proportionality between the athermal resistance to the transformation and the yield strength must be essentially advanced.

{\color{black}
It is rationalized that among three microstructural parameters, domain size, microstrain, and dislocation density, the microstrain plays the dominant role in EOS.  The difference in EOS for
$\alpha$ Zr for the cold-rolled and annealed samples is connected to different microstrains, i.e., internal strains and stresses with significant non-hydrostatic components. It is well known that the EOS under hydrostatic and non-hydrostatic loadings are different, which explains the difference in the EOS for
$\alpha$ Zr for the cold-rolled and annealed samples.  Scatter in data in Figs. \ref{fig:F-falpha}, \ref{fig:F-fomega},  \ref{fig:Ko-kpalpha}, and \ref{fig:Ko-kpomega} is, in particular, due to evolving microstructure and contributions of the defects in producing non-hydrostatic stresses, which affect EOS.   The larger scatter for $\omega$ Zr is because of very different microstructural parameters during pressure increase and decrease.
 A simple model for the initiation of the phase transformation involving microstrain and its concentration is suggested, and a possible model for the growth is outlined.

During transformation, the first rule in the field was found: The average domain size, microstrain, and dislocation density in  $\omega$ Zr for $c<0.8$ are functions of the volume fraction of $\omega$ Zr only, which
  are independent of the plastic strain tensor prior to transformation and pressure.
Some nontrivial patterns in the evolution of microstructural parameters for both phases and cases we found as well.
Since there is a significant jump in dislocation density and domain size from $\alpha$ to $\omega$ Zr, the microstructure is not inherited during phase transformation.

The nontrivial microstructure evolution, different for the pre-deformed and annealed samples, was also found during pressure release.
In particular, a paradoxical result was found:
for the annealed sample, the final dislocation density and average microstrain in the $\omega$ phase are larger than for the severely pre-deformed sample. The following practical suggestion is made for the pressure treatment and irreversible phase transformation: if the goal is to increase the final dislocation density and average microstrain,
it should be done on the annealed sample; to reduce the domain size, the treatment should be applied to the severely pre-strained sample.
 The significant evolution of the microstructure and its effect on phase transformation demonstrates that their postmortem evaluation does not represent the actual conditions during loading.
 }

The obtained results initiate the in situ experimental basis for future predictive structural models for the combined pressure-induced phase transformations and microstructure evolutions.  Comparison with results in \cite{Kumaretal-Acta-20} shows that the effect of pre-straining may be non-monotonous, and more structural states between annealed and maximally hardened
should be studied for the same composition.

\section{Acknowledgments}
Supports of NSF {(DMR-2246991 and CMMI-1943710)} and the ISU (Vance Coffman Faculty Chair  Professorship and Murray Harpole Chair in Engineering) are gratefully acknowledged. We thank Drs. M. T. P\'erez-Prado  and A. P. Zhilyaev
for supplying us with commercially pure Zr purchased from Haines and Maassen (Bonn, Germany) that they studied.
XRD measurements were performed at HPCAT (Sector 16), APS, Argonne National Laboratory. HPCAT operations are supported by the DOE-NNSA Office of Experimental Sciences.  The Advanced Photon Source is a U.S. Department of Energy (DOE) Office of Science User Facility operated for the DOE Office of Science by Argonne National Laboratory under Contract No. DE-AC02-06CH11357.
\begin{appendix}
\section{Microstructural analysis}
\label{microstructureanalysis}
Microstructural analysis was carried out using the Modified Rietveld technique \cite{Rietveld-1969,Young-1993,Mukherjee-JNM-2009} on the XRD patterns for each sample and each pressure step of high-pressure experiments. GSAS-II and MAUD software were used for the refinement of the crystal structure as well as microstructural parameters. First instrumental parameters viz. X-ray wavelength, sample detector distance, and instrumental broadening parameters were obtained using the XRD pattern of the NIST standard $CeO_2$ sample. Subsequently, microstructural parameters viz. average grain/domain size and microstrain and under isotropic size and strain model were refined for each phase of Zr at each pressure step. Using the obtained domain size and microstrain, dislocation density was estimated as
\begin{equation}
\rho_d= (\rho_D\rho_S)^{1/2},
\label{eqn:rho}
\end{equation}
where $\rho_D$ and $\rho_S$ are contributions to dislocation density due to domain size and microstrain respectively \cite{Williamson-1956}. Parameter $\rho_D$ is estimated as
\begin{equation}
\rho_D= 3/D^2,
\label{eqn:rhoD}
\end{equation}
where $D$ is the domain size as estimated from Rietveld refinement. Parameter $\rho_S$ is estimated as
\begin{equation}
\rho_S= k(\varepsilon_l)^2/b^2,
\label{eqn:rhos}
\end{equation}
where $k$ is material constant and given as $6\pi EA/(\mu$  $ln(r/r_\circ))$, $\varepsilon_l$ is lattice strain estimated to be same as average microstrain, $b$ is the modulus of the Burger's vector, $E$ and $\mu$ are Young's modulus and shear modulus, $r$ is the radius of the crystal containing the dislocation, $r_\circ$ is a suitably chosen integration limit \cite{Williamson-1956}, $A$ is a factor depending on the shape of strain distribution and lies between the two extremes of a Cauchy ($\sim 2$) and Gaussian distribution ($\sim \pi/2$).
For $\alpha$ and $\omega$ Zr, moduli  $E$ and $\mu$ and their pressure dependence has been taken from \cite{Fisher-1970} and \cite{Liu-2007}. A reasonable value for $ln(r/r_\circ)$  has been taken as 4 \cite{Williamson-1956}.

For hexagonal systems, three different major slip systems are related to the three glide planes: basal, prismatic, and pyramidal \cite{Lutjering-2002}. When taking into consideration different slip directions and the character of dislocations (edge and screw) in hcp
crystals, there are eleven sub-slip-systems to consider. Generally, more than one sub-slip system is activated during the plastic deformation of materials. Dislocations may also be populated in more than one slip system. For hexagonal systems, most populated  ( $\sim80 \%$) is the basal slip system of type $<a>$ Burger's vector $1/3<11\bar{2}0>$ \cite{Ungar-2002,Takashi-2010, Amos-2020}.  However for $\alpha$  Zr most dominant slip system are prismatic slip system $\{1\bar{1}00\}<11\bar{2}0>$ \cite{Tome-2009,Brown-2006,Tome-2001,Pollock-2013}, again with Burger's vector $1/3<11\bar{2}0>$. For $\omega$ Zr, prismatic $\{11\bar{2}0\}<10\bar{1}0>$ and basal slip system $\{0001\}<10\bar{1}0>$ are the dominant slip system with Burger's vector $1/3<10\bar{1}0>$\cite{Wenk-2013}. Hence using Burger's vectors for the most populated dislocations estimated using pressure-dependent lattice constants, overall dislocation density for each phase of Zr at each pressure step was obtained using equations (\ref{eqn:rho}), (\ref{eqn:rhoD}), and (\ref{eqn:rhos}). The obtained dislocation densities may not be the most accurate, but the trend of evolution with pressure and across the phase transformation in Zr is expected to be the same even if it is estimated using other line profile fitting or whole powder pattern fitting methods.
{\color{black}

\begin{figure}
     \centering
     \begin{subfigure}[b]{0.49\linewidth}
         \centering
         \includegraphics[width=\textwidth]{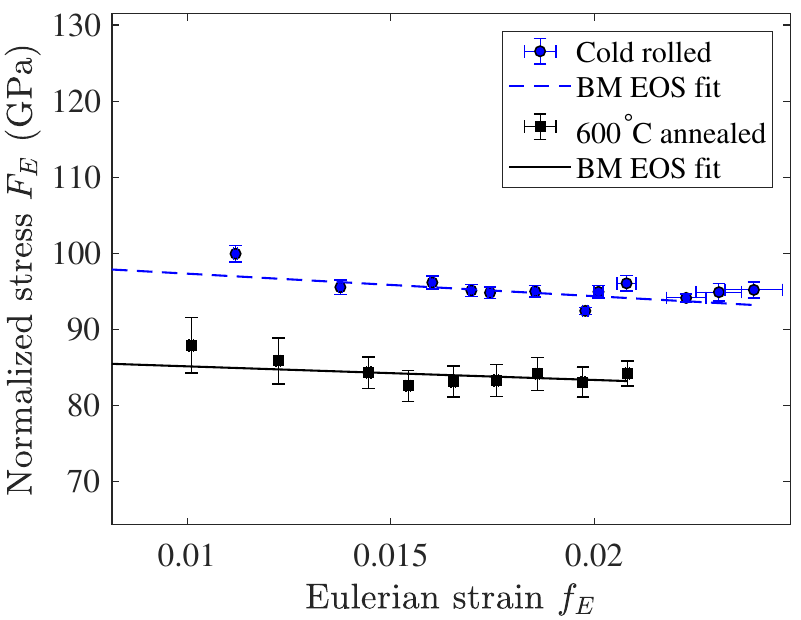}
         \caption{}
         \label{fig:F-falpha}
     \end{subfigure}
     \hfill
     \begin{subfigure}[b]{0.49\linewidth}
         \centering
         \includegraphics[width=\textwidth]{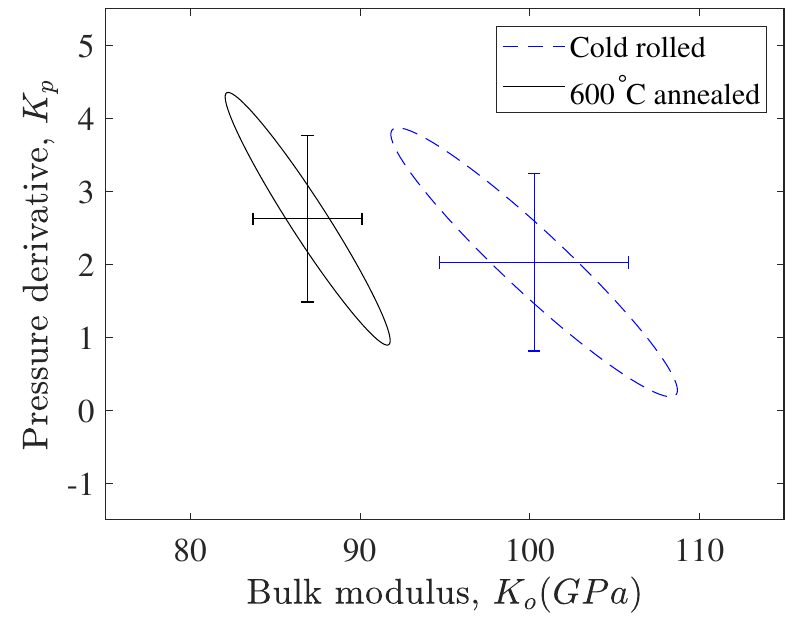}
         \caption{}
         \label{fig:Ko-kpalpha}
     \end{subfigure}

        \caption{Comparison of the $F-f$ plot (a) and the $K_o-K_p$ confidence ellipse plot (b) for 3$^{rd}$-order Birch-Murnaghan equation of state fitting for the $\alpha$ phase of $600^\circ C$ annealed  and cold rolled Zr sample.}

        \label{fig:FfKokplotalpha}
\end{figure}

\begin{figure}
     \centering
     \begin{subfigure}[b]{0.49\linewidth}
         \centering
         \includegraphics[width=\textwidth]{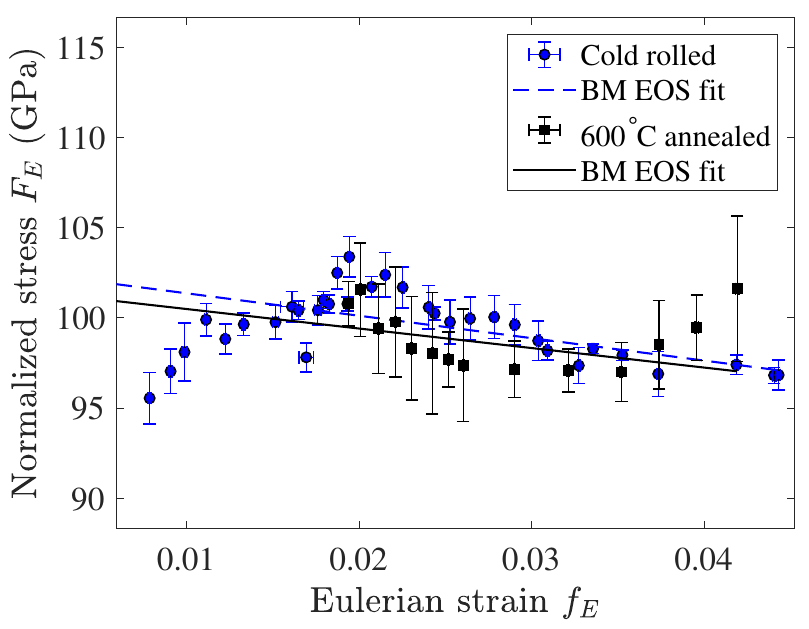}
         \caption{}
         \label{fig:F-fomega}
     \end{subfigure}
     \hfill
     \begin{subfigure}[b]{0.49\linewidth}
         \centering
         \includegraphics[width=\textwidth]{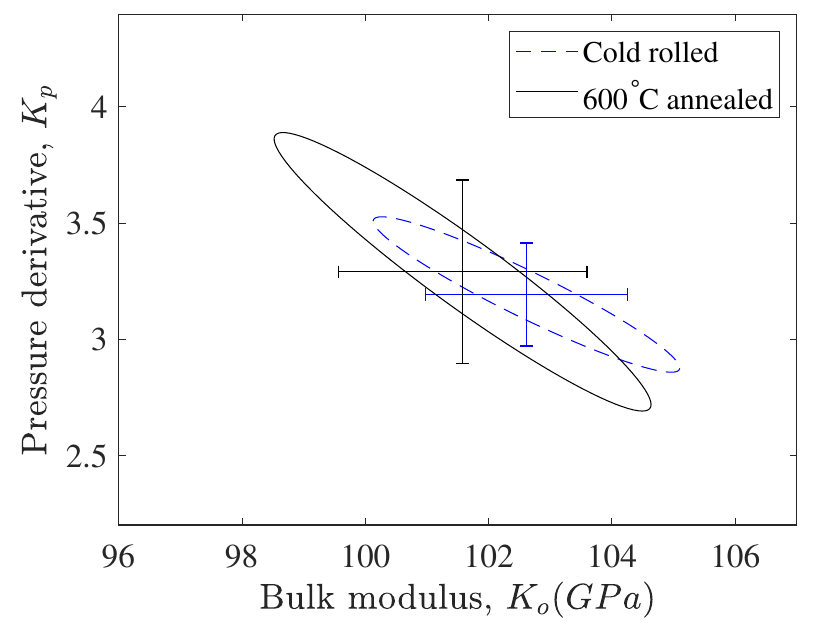}
         \caption{}
         \label{fig:Ko-kpomega}
     \end{subfigure}

        \caption{Comparison of the $F-f$ plot (a) and the $K_o-K_p$ confidence ellipse plot (b) for 3$^{rd}$-order Birch-Murnaghan equation of state fitting for the $\omega$ phase of $600^\circ C$ annealed  and cold rolled Zr sample.}

        \label{fig:FfKokplotomega}
\end{figure}

\section{Equation of state fitting}
\label{BMeosfitting}
The $p-V$ data for both $\alpha$ and $\omega$ phases obtained from analyzing the XRD data was fitted to the well-known 3$^{rd}$-order Birch-Murnaghan equation of state (BM-EOS) \cite{birch-1947} using EOSFIT7-GUI software \cite{Rangel-2016} to obtain bulk modulus and its pressure derivative for both phases as mentioned in the Table \ref{BMfitresults}. For the $\alpha$ phase, only increasing-pressure $p-V$ data was used as the phase transformation was irreversible. For the $\omega$ phase both up-stroke and down-stroke $p-V$ data were used for fitting. Mostly single-phase data points or data points where phase concentration is substantial have been used. The data points where the phase fraction of the concerned phase is small and the Rietveld refinement does not give reliable lattice parameter values have been excluded. Usually, we use single-phase data only to determine an EOS, and after extrapolating it in a two-phase region, we can obtain different pressures in each phase due to a jump in volume and elastic moduli. However, we found here that the pressure difference in phases is negligible, which allowed us to use some points from the two-phase region.

The quality of fitting and the validity of using 3$^{rd}$-order BM-EOS can be illustrated by the plot between Eulerian strain $f=0.5[(V_o/V)^{2/3}-1]$ and normalized stress $F=p/[3f(1+2f)^{5/2}]$ \cite{Jeanloz-1984, Errandonea-2013}, also known as the $F-f$ plot. The $F-f$ plot for $\alpha$ and $\omega$ phases for both $600^\circ C$ annealed and cold rolled samples are compared in Figs. \ref{fig:F-falpha} and \ref{fig:F-fomega}. Corresponding $68.3 \% (1\sigma)$ confidence ellipse plots between bulk modulus ($K_o$) and its pressure derivative ($K_p$) are shown in Figs. \ref{fig:Ko-kpalpha} and \ref{fig:Ko-kpomega}. Scatter in data in Figs. \ref{fig:F-falpha}, \ref{fig:F-fomega},  \ref{fig:Ko-kpalpha}, and \ref{fig:Ko-kpomega} is, in particular, due to evolving microstructure and contributions of the defects in producing non-hydrostatic stresses, which affect EOS.
The data points are more closely spaced in the region of phase transformation and pressure steps are larger in the beginning as finding EOS was not the primary objective during experiments. So for the $\alpha$ phase number of data points is not quite large resulting in relatively less constrained fitting for the $\alpha$ phase. However, there is a discernible difference between the fittings and the obtained bulk modulus and its pressure derivative for the $\alpha$ phase for the cold rolled and the annealed samples. This implies the important role of initial microstructure even for the equation of state. For the $\omega$ phase, combining increasing and decreasing pressure data, the number of data points is large so results are relatively more constrained. However, the larger scatter for the $\omega$ phase is because of very different microstructural parameters during pressure increase and decrease.  Also, the obtained bulk modulus and its pressure derivative for the $\omega$ phase are close for both samples. It is noteworthy that microstructural evolution in the $\omega$ phase, especially microstrain, is similar in both samples. Hence these observations again corroborate the inference of the important role of microstructure in the equation of state of materials.
}
\end{appendix}


\begin{thebibliography}{0}%
\makeatletter
\providecommand \@ifxundefined [1]{%
 \@ifx{#1\undefined}
}%
\providecommand \@ifnum [1]{%
 \ifnum #1\expandafter \@firstoftwo
 \else \expandafter \@secondoftwo
 \fi
}%
\providecommand \@ifx [1]{%
 \ifx #1\expandafter \@firstoftwo
 \else \expandafter \@secondoftwo
 \fi
}%
\providecommand \natexlab [1]{#1}%
\providecommand \enquote  [1]{``#1''}%
\providecommand \bibnamefont  [1]{#1}%
\providecommand \bibfnamefont [1]{#1}%
\providecommand \citenamefont [1]{#1}%
\providecommand \href@noop [0]{\@secondoftwo}%
\providecommand \href [0]{\begingroup \@sanitize@url \@href}%
\providecommand \@href[1]{\@@startlink{#1}\@@href}%
\providecommand \@@href[1]{\endgroup#1\@@endlink}%
\providecommand \@sanitize@url [0]{\catcode `\\12\catcode `\$12\catcode
  `\&12\catcode `\#12\catcode `\^12\catcode `\_12\catcode `\%12\relax}%
\providecommand \@@startlink[1]{}%
\providecommand \@@endlink[0]{}%
\providecommand \url  [0]{\begingroup\@sanitize@url \@url }%
\providecommand \@url [1]{\endgroup\@href {#1}{\urlprefix }}%
\providecommand \urlprefix  [0]{URL }%
\providecommand \Eprint [0]{\href }%
\providecommand \doibase [0]{https://doi.org/}%
\providecommand \selectlanguage [0]{\@gobble}%
\providecommand \bibinfo  [0]{\@secondoftwo}%
\providecommand \bibfield  [0]{\@secondoftwo}%
\providecommand \translation [1]{[#1]}%
\providecommand \BibitemOpen [0]{}%
\providecommand \bibitemStop [0]{}%
\providecommand \bibitemNoStop [0]{.\EOS\space}%
\providecommand \EOS [0]{\spacefactor3000\relax}%
\providecommand \BibitemShut  [1]{\csname bibitem#1\endcsname}%
\let\auto@bib@innerbib\@empty
\end{thebibliography}%


\begin{thebibliography}{00}
\bibitem{Bridgman-52}
Bridgman P. W. Studies in Large Plastic Flow and Fracture, McGraw-
Hill, New York (1952).

 \bibitem{lubliner-1990}
 Lubliner J.
 Plasticity Theory. New York, Macmillan Publishing Company (1990).

 \bibitem{levitas-book96}
 Levitas V.I.
 Large Deformation of Materials with
 Complex Rheological Properties at Normal and High
 Pressure. New York, Nova Science Publishers (1996).


\bibitem{edalati2022nanomaterials}
Edalati K.,  Bachmaier A.,  Beloshenko V.,  Beygelzimer Y.,  Blank V.,  Botta W.,  Bry{\l}a K., \v{C}í\v{z}ek J.,  Divinski S.,  Enikeev N.,  Estrin Y.,  Faraji G.,  Figueiredo B.,  Fuji M.,  Furuta T.,  Grosdidier T.,  Gubicza J.,  Hohenwarter A.,  Horita Z.,  Huot J.,  Ikoma Y.,  Jane\v{c}ek M.,  Kawasaki M.,  Kr\v{a}l P.,  Kuramoto S.,  Langdon T.,  Leiva D.,  Levitas V.I.,  Mazilkin A.,  Mito M.,  Miyamoto M.,  Nishizaki T.,  Pippan R.,  Popov V.,  Popova E.,  Purcek G.,  Renk O., Révész  Á.,  Sauvage X.,  Sklenicka V.,  Skrotzki W.,  Straumal B.,  Suwas S.,  Toth L.,  Tsuji N.,  Valiev R.,  Wilde  G.,  Zehetbauer M.,  Zhu X.
     Nanomaterials by severe plastic deformation: review of historical developments and recent advances. { Materials Research Letters}, \textbf{10}, 163-256 (2022).


\bibitem{levitas-mechchem-04} 
V. I. Levitas, Continuum Mechanical Fundamentals of Mechanochemistry, in {\it High Pressure Surface Science and Engineering}, 159-292 (Institute of Physics, Bristol, 2004).
\bibitem{levitas-prb-04}
V. I. Levitas, High-Pressure Mechanochemistry: Conceptual Multiscale  Theory and Interpretation of Experiments, Phys. Rev. B {\bf 70}, 184118 (2004).
\bibitem{Blank-Estrin-2014}
V. D. Blank and E. I. Estrin, {\it Phase Transitions in Solids under High Pressure} (CRC Press, Boca Raton, 2014).
\bibitem{Edalati-Horita-16}
K. Edalati and  Z. Horita, A review on high-pressure torsion (HPT) from 1935 to 1988, Mat. Sci. Eng. A. {\bf 652}, 325-352 (2016).
 \bibitem{Gaoetal-19}
Y. Gao, Y. Ma, Q. An, V. I. Levitas, Y. Zhang, B. Feng, J. Chaudhuri, W. A. Goddard III,  Shear strain driven formation of nano-diamonds at sub-gigapascals and 300 K, Carbon
{\bf 146}, 364-368 (2019).
\bibitem{Levitas-JPCM-18}
V.I. Levitas, High pressure phase transformations revisited.
  Journal of Physics: Condensed Matter, {\bf 30}, 163001 (2018).
\bibitem{Levitas-MT-19}
V.I.  Levitas, High-Pressure Phase Transformations under Severe Plastic Deformation by Torsion in Rotational Anvils. Material Transactions, {\bf 60}, 1294-1301 (2019).
\bibitem{Levitas-MT-23}
Levitas V.I. Recent in situ Experimental and Theoretical Advances in Severe Plastic Deformations, Strain-Induced Phase Transformations, and Microstructure Evolution under High Pressure.  Material Transactions, {\bf 64}, 1866-1878 (2023).


\bibitem{Zilbershtein-75}
 V. A. Zilbershtein, N. P. Chistotina, A. A. Zharov, N. S. Grishina, and E. I. Estrin, Alpha-omega transformation in titanium and zirconium during shear deformation under pressure, Fizika Metallov I Metallovedenie {\bf 39}, 445–447 (1975).
\bibitem{Banerjee-2007}
S. Banerjee and P. Mukhopadhyay,
{\it Phase Transformations: Examples from Titanium and Zirconium Alloys, Pergamon Materials Series, 1st edition} (Elsevier, Amsterdam, 2007).
\bibitem{Perez-2009}
M. T. P\'erez-Prado  and A. P. Zhilyaev, First experimental observation of shear induced hcp to bcc transformation in pure Zr, Phys. Rev. Lett. {\bf  102}, 175504 (2009).
\bibitem{Edalatietal-MSEA-2009}
K. Edalati, Z. Horita, S. Yagi, and E. Matsubara, Allotropic phase transformation of pure zirconium by high-pressure torsion, Mat. Sci. Eng. A {\bf 523},  277–281 (2009).

\bibitem{Zhilyaevetal-MSEA-2011}
A. P. Zhilyaev, I. Sabirov, G. Gonz\'alez-Doncel, J. Molina-Aldaregu\'ia, B. Srinivasarao, and M. T. P\'erez-Prado,  Effect of Nb additions on the microstructure, thermal stability and mechanical behavior of high pressure Zr phases under ambient conditions, Mat. Sci. Eng. A {\bf 528}, 3496–3505 (2011).

\bibitem{KKP-Levitas-Acta-2020}
K. K. Pandey, V. I. Levitas, In situ quantitative study of plastic strain-induced phase transformations under high pressure: Example for ultra-pure Zr, Acta Materialia, {\bf 196}, 338-346 (2020).
\bibitem{Lin-Levitas-MRL-23}
Lin F.,  Levitas V.I.,  Pandey K.K.,  Yesudhas S., and
  Park C. In-situ study of rules of nanostructure evolution, severe plastic deformations, and friction under high pressure.
Materials Research Letters, {\bf  11}, 757-763 (2023).

\bibitem{Velisavljevic-etal-11}
Velisavljevic N., Chesnut G. N., Stevens L. L., and
 Dattelbaum D. M. Effects of interstitial impurities on the high pressure martensitic $\alpha$ to $\omega$ structural
transformation and grain growth in zirconium.  J. Phys.: Condens. Matter, {\bf 23}, 125402 (2011).

\bibitem{Liu-etal-23}
Liu L., Jing Q., Geng H.Y., Li Y., Zhang Y., Li J., Li S., Chen X., Gao J., Wu Q. Revisiting the
High-Pressure Behaviors of Zirconium: Nonhydrostaticity
Promoting the Phase Transitions and Absence of the Isostructural Phase Transition in b-Zirconium. Materials, {\bf 16}, 5157 (2023).

\bibitem{Anzellini-etal-20}
Anzellini S.,  Bottin F., Bouchet J., and Dewaele A.  Phase transitions and equation of state of zirconium under high pressure.
Phys. Rev. B, {\bf 102}, {184105} (2020).

\bibitem{Kumaretal-Acta-20}
M. A. Kumar,  N. Hilairet, R.J. McCabe, T. Yu, Y. Wang, I.J. Beyerlein, C.N. Tom\`{e},
Role of twinning on the omega-phase transformation and stability in zirconium.
Acta Materialia {\bf 185},  211–217 (2020).


\bibitem{Levitas-Zarechnyy-PRB-DAC-10}
V. I. Levitas, O. Zarechnyy,    Modeling and simulation of strain-induced phase transformations under compression in a diamond anvil cell. Phys. Rev. B., {\bf 82}, 174123 (2010).


\bibitem{Levitas-Zarechnyy-PRB-RDAC-10}
V. I. Levitas, O. Zarechnyy,  Modeling and simulation of strain-induced phase transformations under compression and torsion in a rotational diamond anvil cell. Phys. Rev. B. {\bf 82}, 174124 (2010).
\bibitem{Feng-Levitas-IJP-BN-RDAC-19}
B. Feng, V. I.  Levitas,  W. Li,  FEM modeling of plastic flow and strain-induced phase transformation in BN under high pressure and large shear in a rotational diamond anvil cell. International Journal of  Plasticity, {\bf 113}, 236-254 (2019).





\bibitem{Levitas-JMPS-I-97}
V.I. Levitas,
 Phase Transitions in Elastoplastic Materials: Continuum
 Thermomechanical Theory and Examples of Control. Part I.
J. Mech. Phys. Solids, {\bf  45}, 923-947 (1997).


\bibitem{Levitas-JMPS-II-97}
V.I. Levitas,
 Phase Transitions in Elastoplastic Materials: Continuum
 Thermomechanical Theory and Examples of Control. Part  II.
J. Mech. Phys. Solids, {\bf 45},  1203-1222 (1997).


\bibitem{levitas-ijss1998}
V.I. Levitas,
 Thermomechanical Theory of Martensitic Phase Transformations
 in Inelastic Materials.
Int. J. Solids and Structures, {\bf 35}, 889-940 (1998).  


\bibitem{Levitas-IJP-21}
 Levitas V.I. Phase transformations, fracture, and other structural changes in inelastic materials.
International Journal of Plasticity, {\bf 140}, 102914 (2021).

\bibitem{Levitasetal-PhilMag-02}
V.I. Levitas, A.V. Idesman, G.B. Olson, E. Stein,
 Numerical Modeling of Martensite Growth in Elastoplastic Material.
Philosophical Magazine, {\bf A82}, 429-462 (2002). 



\bibitem{olson+cohen-1986}
G.B. Olson, M. Cohen,  Dislocation Theory of Martensitic Transformations. in {\it
F R N Nabarro (Ed.)  Dislocations in Solids} {\bf 7} , 297-407  (Elsevier Science Publishers BV 1986).




\bibitem{Xu-Khachaturyan-etal-NPJ-CM-18}
Y.C. Xu,  W.F. Rao,  J.W. Morris, A.G. Khachaturyan,   Nanoembryonic thermoelastic equilibrium and enhanced properties of defected pretransitional materials, npj Computational Materials, {\bf 4}, 58 (2018).


\bibitem{Xu-Khachaturyan-etal-AM-19}
Y.C. Xu,   C. Hu,  L. Liu,  J. Wang, W.F. Rao,  J.W. Morris, A.G. Khachaturyan,  A nano-embryonic mechanism for superelasticity, elastic softening, invar and elinvar effects in defected pre-transitional materials. Acta. Mater., {\bf 171}, 240-252 (2019).

\bibitem{levitas-javanbakht-APL-13}
V. I. Levitas, M. Javanbakht, Phase field approach to interaction of phase transformation and dislocation evolution, Appl. Phys. Lett., {\bf 102}, 251904 (2013).

\bibitem{Javanbakht-Levitas-JMS-18}
M. Javanbakht, V.I. Levitas,
Nanoscale mechanisms for high-pressure mechanochemistry: a phase field study.
 Journal of Materials Science, {\bf 53}, 13343-13363 (2018).





\bibitem{Vickers}
R.L. Smith and G.E. Sandland, An Accurate Method of Determining the Hardness of Metals, with Particular Reference to Those of a High Degree of Hardness, Proceedings of the Institution of Mechanical Engineers, {\bf 1}, 623–641 (1922).


\bibitem{fit2d1}
A. P. Hammersley, Fit2d: An introduction and overview,  in {\it ESRF Internal Report,
ESRF97HA02T} (Institute of Physics, Bristol, 1997).
\bibitem{fit2d2}
A. P. Hammersley, S. O. Svensson, M. Hanfland, A. N. Fitch, and D. Hausermann, Two-dimensional detector software: From real detector to idealized image or two-theta scan. High Press. Res. {\bf 14}, 235–248 (1996).

\bibitem{gsas2}
B. H. Toby and R. B. Von Dreele, GSAS-II: the genesis of a modern opensource
all purpose crystallography software package, J. Appl. Cryst. {\bf 46}, 544–549 (2013).
\bibitem{MAUD}
M. Ferrari and L. Lutterotti, Method for the simultaneous determination of anisotropic residual stresses and texture by X-ray diffraction, J. Appl. Phys., {\bf 76}, 11, 7246-55 (1994).

\bibitem{Yongrong-RSI2004}
Y. Shen, R.S. Kumar, M. Pravica, and M. F. Nicol, Characteristics of silicone fluid as a pressure transmitting medium in diamond anvil cells, Rev. Sci. Inst. {\bf 75}, 4450 (2004).

\bibitem{Torikachvili-RSI2015}
M. S. Torikachvili, S. K. Kim, E. Colombier, S. L. Bud’ko, and P. C. Canfield, Solidification and loss of hydrostaticity in liquid media, Rev. Sci. Inst. {\bf 86}, 123904 (2015).
used for pressure measurements
\bibitem{Klotz-2009}
S Klotz, J-C Chervin, P Munsch, and G Le Marchand, Hydrostatic limits of 11 pressure
transmitting media, J. Phys. D: Appl. Phys. {\bf 42},  075413 (2009).

\bibitem{Dewaele-PRB-2004}
A. Dewaele, P.  Loubeyre and M. Mezouar, Equations of state of six metals above 94 GPa, Phys. Rev. B {\bf 70}, 094112 (2004).

\bibitem{RDAC-ISM-15}
N. V. Novikov,   L. K. Shvedov,  Yu. N. Krivosheya, V. I. Levitas,
New Automated Shear Cell with Diamond Anvils for in situ Studies of Materials Using X-ray Diffraction, Journal of Superhard Materials,  {\bf 37}, 1,  1-7 (2015).

\bibitem{birch-1947}
F. Birch, Finite Elastic Strain of Cubic Crystals, Physical Review, {\bf 71} (11), 809–824 (1947).

\bibitem{Levitas-PRB-21}
Levitas V.I.    Nonlinear elasticity of prestressed single crystals at high pressure
    and various elastic moduli.  Physical Review B, {\bf 104},  214105 (2021).


\bibitem{Levitas-etal-NatCom-23}
 Levitas V.I.,  Dhar A., and Pandey K.K. Tensorial stress-plastic strain fields in $\alpha$ - $\omega$ Zr mixture, transformation kinetics, and friction in diamond anvil cell.  Nature Communication, {\bf 14}, 5955 (2023).

\bibitem{Rangel-2016}
J. Gonzalez-Platas, M. Alvaro, F. Nestola and R. Angel, EosFit7-GUI: a new graphical user interface for equation of state calculations, analyses and teaching, J. Appl. Cryst. {\bf 49}, 1377-1382 (2016).
\bibitem{Blanketal-InorMat-83}
V. D. Blank, Y. S. Konyaev, V. T. Osipova, E. I. Estrin, Influence of phase hardening and plastic-deformation on the hysteresis of polymorphous transformation in alkali-halide crystals under pressure, Inorganic Materials {\bf 19}, 72-76 (1983).
\bibitem{Popovetal-SciRep-19}
D Popov, N Velisavljevic, W Liu, R Hrubiak, C Park, G Shen, Real time study of grain enlargement in zirconium under room-temperature compression across the $\alpha$ to $\omega$ phase transition
Scientific reports {\bf 9}, 1-7 (2019).
\bibitem{Tome-2009}
L. Capolungo, I. J. Beyerlein, G. C. Kaschner, C. N. Tom\`{e}, On the interaction between slip dislocations and twins in HCP Zr, Mater. Sci. Eng., A {\bf 513}, 42-51 (2009).
\bibitem{Brown-2006}
G. C. Kaschner, C. N. Tom\`{e}, I. J. Beyerlein, S. C. Vogel, D. W. Brown, R. J. McCabe, Role of twinning in the hardening response of zirconium during temperature reloads, Acta Mater. {\bf 54}, 2887-2896 (2006).
\bibitem{Tome-2001}
C. N. Tom\`{e}, P. J. Maudlin, R. A. Lebensohn, G. C. Kaschner, Mechanical response of zirconium-I. Derivation of a polycrystal constitutive law and finite element analysis, Acta Mater. {\bf49}, 3085-3096 (2001).

\bibitem{Pollock-2013}
M. Knezevic, I. J. Beyerlein, T. Nizolek, N. A. Mara, T. M. Pollock, Anomalous Basal Slip Activity in Zirconium under High-strain Deformation, Mater. Res. Lett. {\bf 1}, 133-140 (2013).

\bibitem{Meyers-Chawla-02}
Meyers M.A. and Chawla K.K. Mechanical behavior of materials. Prentice Hall, Upper Saddle River, NJ, USA, 2002.

\bibitem{Rietveld-1969}
H. M. Rietveld, A profile refinement method for nuclear and magnetic structures, J. Appl. Cryst. {\bf 2}, 65–71 (1969).
\bibitem{Young-1993}
R. A. Young, {\it The Rietveld Method}, (International Union of Crystallography, Oxford University Press, 1993).
\bibitem{Mukherjee-JNM-2009}
P. Mukherjee, A. Sarkar, M. Bhattacharya, N. Gayathri, P. Barat, Post-irradiated microstructural characterization of cold-worked SS316L by X-ray diffraction technique, J. Nuclear Materials, {\bf 395}, 37-44 (2009).

\bibitem{Williamson-1956}
G. K. Williamson, R. E. Smallman, Dislocation densities in some annealed and cold-worked metals from measurements on the Xray debye-scherrer spectrum, Philos. Mag. {\bf 1}, 34-46 (1956).

\bibitem{Fisher-1970}
E. S. Fisher, M. H. Manghnani, T. J. Sokolowski, Hydrostatic Pressure Derivatives of the Single-Crystal Elastic Moduli of Zirconium, J. Appl. Phys. {\bf 41}, 2991-2998 (1970).

\bibitem{Liu-2007}
W. Liu, B. Li, L. Wang, J. Zhang, Y. Zhao, Elasticity of $\omega$ -phase zirconium  Phys. Rev. B {\bf 76}, 144107 (2007).


\bibitem{Lutjering-2002}
G. Lutjering, J. Williams, A. Gysler, {\it Microstructures and Mechanical Properties of Titanium Alloys} (World Scientific: Singapore 2002).

\bibitem{Ungar-2002}
I. C. Dragomir, T. Ungar, Contrast factors of dislocations in the hexagonal crystal system, J. Appl. Cryst. {\bf 35}, 556-564 (2002).
\bibitem{Takashi-2010}
Takashi Shintani, Yoshinori Murata, Yoshihiro Terada, Masahiko Morinaga, Evaluation of Dislocation Density in a Mg-Al-Mn-Ca Alloy Determined by X-ray Diffractometry and Transmission Electron Microscopy, Materials Transactions,  {\bf 51},(6), 1067-1071 (2010).
\bibitem{Amos-2020}
Amos Muiruri, Maina Maringa, Willie du Preez, Evaluation of Dislocation Densities in Various
Microstructures of Additively Manufactured Ti6Al4V (Eli) by the Method of X-ray Diffraction, Materials, {\bf 13}, 5355 (2020).
\bibitem{Wenk-2013}
H. R. Wenk, P. Kaercher, W. Kanitpanyacharoen, E. Zepeda-Alarcon, Y. Wang, Orientation Relations During the $\alpha-\omega$ Phase Transition of Zirconium: In Situ Texture Observations at High Pressure and Temperature, Phys. Rev. Lett. {\bf 111}, 195701 (2013).
\bibitem{Jeanloz-1984}
D. L. Heinz and R. Jeanloz, J. Appl. Phys. 55, 885 (1984).
\bibitem{Errandonea-2013}
D. Errandonea, R. S. Kumar, O. Gomis, F. J. Manjon, V. V. Ursaki, and I.
M. Tiginyanu, J. Appl. Phys. 114, 233507 (2013).



%
%
%
%
%
%





\end{thebibliography}
\end{document}